\documentclass[]{elsart}

 \usepackage{graphics}
 \usepackage{epsfig}

\usepackage{amssymb}

\usepackage{subfigure}
\begin{document}

\begin{frontmatter}



\title{Experimental constraints on the astrophysical interpretation of the cosmic ray Galactic-extragalactic transition region}


\author[A]{C. De Donato\corauthref{cor1}},
\ead{cinzia.dedonato@mi.infn.it}
\corauth[cor1]{Corresponding author. Dipartimento di Fisica, Universit\`a degli Studi di Milano, via Celoria 16, Milano, CAP 20133, Italy. Telephone number: ++39 0250317710}
\author[B]{G. A. Medina-Tanco\corauthref{cor2}}
\corauth[cor2]{Corresponding author. Instituto de Ciencias Nucleares, UNAM, Apartado Postal 70-543, Ciudad Universitaria, Mexico D.F. 04510, Mexico. Telephone number: ++52-55-5622-4690 (ext.: 333), fax number: ++52-55-5622-4693}
\ead{gmtanco@nucleares.unam.mx}

\address[A]{Dipartimento di Fisica, Universit\`a degli Studi di Milano and INFN, Milano, Italy}
\address[B]{Dep. Altas Energias, Inst. de Ciencias Nucleares, Universidad Nacional Autonoma de M\'exico, M\'exico DF, CP 04510}

\begin{abstract}
The energy region spanning from $\sim 10^{17}$ to $\lesssim 10^{19}$ eV is
critical for understanding both, the Galactic and the extragalactic cosmic ray
fluxes. This is the region where the propagation regime of nuclei inside the
Galactic magnetic environment changes from diffusive to ballistic, as well as
the region where, very likely, the most powerful Galactic accelerators reach their
maximum output energies.  In this work, a diffusion Galactic model is used to
analyze the end of the Galactic cosmic ray spectrum and its mixing with the
extragalactic cosmic ray flux. In particular, we study the conditions that must
be met, from the spectral and composition points of view, by the Galactic and
the extragalactic fluxes in order to reproduce simultaneously the total spectrum
and elongation rate measured over the transition region by HiRes and Auger.
Our analysis favors a mixed extragalactic spectrum in combination with a Galactic
spectrum enhanced by additional high energy components, i.e., extending beyond
the maximum energies expected from regular supernova remnants. The two
additional components have mixed composition, with the lowest energy one
heavier than the highest energy one. The potential impact on the astrophysical
analysis of the assumed hadronic interaction model is also assessed in detail.

\end{abstract}

\begin{keyword}
cosmic rays \sep Galactic deconfinement \sep composition and spectrum
\PACS 
\end{keyword}
\end{frontmatter}

\addcontentsline{toc}{section}{Introduction}
\section*{Introduction}
The cosmic ray (CR) energy spectrum extends for many orders of magnitude
with a power law  index $\approx2.7$. Along this range of energies,
the three spectral features are known: the first knee at $E\approx
3~PeV$, the second knee at $E\approx0.5~EeV$ and the ankle, a dip
extending from the second knee to beyond $10~EeV$. The nature of the
second knee and of the ankle is still uncertain \cite{GMTanco2006};
a possible interpretation of the two features is the transition
between the Galactic and extragalactic components. At energies
between $10^{17}-10^{18}~eV$ the Galactic supernova remnants (SNR)
 are expected to become inefficient as accelerators. This
fact, combined with magnetic deconfinement, should mark the end of
the Galactic component of cosmic rays, although the picture could be
confused by the existence of additional Galactic accelerators at
higher energy. On the other hand, at energies above the second knee,
extragalactic particles are able to travel from the nearest
extragalactic sources in less than a Hubble time. Consequently, the
spectrum may present above $10^{17.5}~eV$ a growing extragalactic
component that becomes dominant above $10^{19}~eV$. The region
between the second knee and the ankle could be the transition region
between the Galactic and extragalactic components.\\

 In this work we analyze the transition region comparing the diffusive Galactic
spectrum from SNRs with two different models of extragalactic
spectrum, one in which only protons \cite{Berezinsky2006c} are
injected at the sources and another in which a mixed composition
containing heavy nuclei \cite{Allard2007} is injected. 
To discriminate between the  possible  astrophysical scenarios we 
analyze the composition parameter  $X_{max}$  inferred from the different extragalactic (EG) 
models using  the hadronic interaction models  used in literature. 
A comparison with the experimental data is done with special attention to Auger and HiRes data.

\section{Galactic and extragalactic cosmic rays: transition models}

The nature of the knee, consisting of a steepening of the spectrum from a power low index $\gamma_g=2.7$ to  $\gamma_g=3.1$,
is well explained by the rigidity-models (rigidity-acceleration model and rigidity-confinement model), in which the maximum energy achievable 
by nuclei is rigidity dependent. In these models, the knees of the spectrum of nuclei of charge Z are related to the 
proton knee energy $E^Z_{kn}= Z E^p_{kn}$, where $E^p_{kn}=2.5\times10^{15}~eV$. Beyond the highest energy knee $E^{Fe}_{kn}=6.5\times10^{16}eV$ 
the total Galactic flux, dominated by the Iron component, must be steeper. 
While this spectral feature seems to be confirmed by KASCADE data \cite{Kascade2005}, the nature of the
second knee and of the ankle are still uncertain \cite{GMTanco2006}. 
Although the region between the second knee and the ankle is naturally considered as the transition between the Galactic and extragalactic 
component, the way the transition takes place is very model dependent.
The main models describing the transition are the \emph{ankle model}, the \emph{dip model} and the \emph{mixed composition model}. 

In the \emph{ankle model} the transition between the Galactic cosmic rays (GCR) and extragalactic cosmic rays (EGCR) 
 occurs at the ankle $E_a\approx1\times 10^{19}~eV$ \cite{BiermannGMT2003,Hillas2006,Hillas2005,Hillas2004,Wibig2005,Waxman1995,Vietri1995,Hill1985}. 
The dip in the CR spectrum is explained as the intersection of a flat extragalactic component with the steep Galactic component. 
An advantage of this model is the extragalactic flat generation spectrum $E^{-2}$ which provides reasonable luminosities of the sources. 
On the other hand, the \emph{ankle model} doesn't work well in the framework of the rigidity-model. 
If the transition from Galactic to extragalactic CRs occurs at the ankle, other additional Galactic  mechanisms are required 
to accelerate particles at energies beyond $E^{Fe}_{kn}=6.5\times10^{16}eV$.

A different explication of the dip and a lower transition energy comes from the \emph{dip model} \cite{Berezinsky2002,Berezinsky2006c}. 
In this model the dip is caused by $e^+e^-$ pair production by the extragalactic protons with the CMB photons. 
The transition energy is determined by the energy at which adiabatic energy losses are equal to pair production energy losses. 
In this case the transition between the Galactic component and the extragalactic 
component takes place  at lower energies, around the second knee, in agreement with the rigidity-model. 
The calculated position and shape  of the dip, confirmed by Akeno-AGASA \cite{AGASA2003}, HiRes \cite{HiRes2004}, 
Yakutsk \cite{Yakutsk2003,Yakutsk2004} and Fly's Eye \cite{FlysEye1994} experimental data\begin{footnote}{Auger data 
don't contradict the high energy part of dip but an extension of Auger to lower energies it's needed to confirm 
the feature of the dip \cite{AugerFD_ICRC2007}.}\end{footnote},
is ``universal'' as it doesn't depend on the type of propagation, 
on the source density and separation (for separation distance $d<100~Mpc$) and on the evolutionary model 
of the sources. The only factor that affect the dip is a heavy component of the EG spectrum \cite{Berezinsky2007a,Aloisio2007c}. 
The shape and position of the dip is in agreement with observations for an EG pure proton composition with a maximum contamination of Helium nuclei of the order of $10\%$.\\
On the other hand, as the dip model requires a generation spectrum $E^{-2.7}$, 
a solution is needed to prevent the too high emissivity 
needed  at lower energies \cite{Berezinsky2002,Berezinsky2004,Berezinsky2006c,Kachelriess2006}.

An alternative and intermediate model is the \emph{mixed composition model} \cite{Allard2005a,Allard2005b,Allard2007}. 
In this model the EG cosmic ray composition is assumed to be mixed, 
in analogy with the Galactic component. As in the \emph{ankle model}, the intersection of the Galactic and extragalactic 
component gives origin to the the dip structure but with the advantage of a lower transition energy (around $E\approx3\times 10^{18}~eV$), which 
softens the requirement of additional acceleration mechanism. 
The calculated spectrum, fitted with the observed data,  
presents a spectral index $\gamma\approx 2.1-2.3$, which also provides a reasonable luminosity. 

The three model can be experimentally distinguished through measurements of the spectrum and of anisotropies, although the 
most discriminant  feature is the chemical composition.
While in the ankle model the Galactic heavy component dominates up to the ankle energy, in the dip model the transition is completed at energy around $1\times 10^{18}~eV$ where the composition is proton dominated. 
The composition at this energy (Iron/proton) is a discriminant between the two models. 
In the case of the mixed model, the transition occurs at $E\approx3\times 10^{18}~eV$ and the chemical composition in the dip region is mixed. 
This model predicts a slower decrease of the Fe component and a slower increase of the proton fraction in the transition energy range.

\section{Galactic-extragalactic spectrum}

The current paradigm for Galactic cosmic ray acceleration is the Fermi acceleration mechanism by shock waves of 
 SuperNova Remnants (SNRs) \cite{Biermann2003}.\\
In this section we calculate the Galactic diffusive spectrum from SNRs 
using the numerical diffusive propagation code GALPROP \cite{Strong1998,Strong2001}.\\
The calculated Galactic spectrum from SNRs is then combined with two different models of EG spectrum, 
one in which only protons \cite{Berezinsky2006c} are injected at the sources and another in which a mixed composition
containing heavy nuclei \cite{Allard2007} is injected instead.\\
We analyze the transition region between Galactic and extragalactic components in the two different EG scenarios 
and compare the combined total spectrum with the available experimental data.

\subsection{Diffusion Galactic model}\label{DiffModel}

We use the numerical diffusive propagation code
GALPROP \cite{Strong1998,Strong2001} to reproduce the Galactic
spectrum from SNRs.

The diffusive model is axisymmetric. The propagation region is, in cylindrical coordinates,
bounded by $R=R_h=30~kpc$ and $z=z_h=4~kpc$, beyond which free
escape is assumed. The diffusion equation is:

\vskip -0.3cm

\begin{equation}\label{eq:diffeq}
\frac{\partial\psi}{\partial t}  =  q(\vec{r},p)+ \vec{\nabla}
\times (D_{xx}\vec{\nabla} \psi)  -\frac{\partial}{\partial
p}(\dot{p}\psi)-\frac{1}{\tau_f}\psi-\frac{1}{\tau_r}\psi
\end{equation}

where $\psi(\vec{r},p,t)$ is the density per unit of total particle
momentum, $q(\vec{r},p)$ is the source term, $D_{xx}$ is the spatial
diffusion coefficient, $\dot{p}= dp/dt$ is the momentum loss rate
and $\tau_f$ and $\tau_r$ are the time scale of fragmentation and
the time scale of radioactive decay respectively.

The GALPROP code solves the diffusion equation for all cosmic-ray species 
starting from the heaviest nucleus and then proceeding to lighter nuclei 
using the computed secondary source functions. 
The numerical solution of the transport equation is based on 
a Crank-Nicholson \cite{Press1992,Strong1998} implicit second-order scheme.

The diffusion coefficient is taken as $\beta D_0(\rho / \rho_D)^{\delta}$, where
$\rho$ is the particle rigidity, $D_0$ is the diffusion coefficient 
at a reference rigidity $\rho_D$ and $\delta=0.6$. 
 The diffusion coefficient can be inferred from the abundances of light nuclei, produced mainly 
through spallation of heavy elements, as Li, Be and B, which give an estimation of the 
 time of residence of CRs in the galaxy of $\approx1.5\times10^7~yr$ \cite{Simpson1988}. 
We use $D_0=5.75\times 10^{28}cm^2 s^{-1}$ at the reference rigidity  $\rho_{D}=4~GV$.\\

The assumption of a diffusion coefficient with an energy dependence $E^{-0.6}$
is not universally accepted \cite{Biermann1993}. In fact, the turbulence in
the interstellar medium seems to follow closely a Kolmogorov spectrum, which
should lead to an energy dependence of $E^{1/3}$ for the diffusion coefficient.
In this scenario, the difference between primary and secondary cosmic ray 
energy spectra could be explained by the trapping of primary cosmic rays 
in high density regions with Kolmogorov turbulence near their acceleration 
sites, where secondary CR would be mainly produced, and the diffusion of the 
latter inside this region and the interstellar medium. However, since at high
energy both models produce similar spectra and composition profiles, we think
our analysis is to a large extent independent of these assumptions.

The assumed distribution of cosmic rays sources is the one reproducing the cosmic-ray 
distribution determined by the analysis of EGRET gamma-ray data, which has the same parameterization of the R-dependence as that used for SNRs \cite{Strong1998}:

\begin{equation}\label{eq:Source_distr}
f(R,z)= f(R)\exp{(-|z|/z_{scale})}
\end{equation}

\noindent where $z_{scale} = 0.2$ kpc and: 

\begin{displaymath}
f(R)= \left\{ \begin{array}{ll}
\label{fr}
 19.3 &~~~ R\leq4~kpc,\\
 21.9 &~~~ 4~kpc<R\leq8~kpc,\\
 15.8 &~~~ 8~kpc<R\leq10~kpc,\\
 18.3 &~~~ 10~kpc<R\leq12~kpc,\\
 13.3 &~~~ 12~kpc<R\leq15~kpc,\\
 7.4 &~~~ R>15~kpc.
\end{array}\right.
\end{displaymath}

This choice is due to the fact that the SNR distribution \cite{CaseBhattacharya1996} produces CR distribution distinctly different from the measurements.
A solution to the apparent discrepancy between the radial gradient in the diffuse Galactic 
gamma-ray emissivity and the distribution of SNRs  was proposed by Strong et al. \cite{Strong2004}.

According to shock acceleration models \cite{Gaisser1991,Protheroe2004a},
the injection spectrum is a power law function in rigidity with a break at rigidity $\rho_0$, beyond which it falls exponentially with a rigidity scale $\rho_c$:

\begin{displaymath}
I(\rho)= \left\{ \begin{array}{ll}
\label{InjSp}
 \left(\frac{\rho}{\rho_0}\right)^{-\alpha} & ~~~\rho<\rho_0,\\
\exp\left[-\left(\frac{\rho}{\rho_0}-1\right)/\frac{\rho_c}{\rho_0})\right] & ~~~\rho>\rho_0, \nonumber
\end{array}\right.
\end{displaymath}

where $\rho_0=1.8~PV$, $\rho_c=1.26~PV$ and $\alpha=2.05$, 
which is the case of strong shock waves ($M>>1$) \cite{Krymskii1977,Blandford1978}.

Stable nuclei with $Z\leq26$ are injected, with energy independent isotopic
abundances derived from low energy CR measurements  
\cite{Strong2001,Hesse1996,Duvernois1996a,Duvernois1996b,Duvernois1996c,Thayer1997,Connel1997}.

The interstellar hydrogen distribution, molecular, atomic and ionized (H2, HI, HII), is derived from radio
 HI and CO surveys in 9 Galactocentric rings and  information on the ionized component \cite{Moskalenko2002}.
The distribution of molecular hydrogen is derived indirectly from CO radio-emission and the assumption that the conversion factor H2 /CO is the same for the whole Galaxy \cite{StrongMattox1996}.  
The atomic hydrogen (HI) distribution is taken from \cite{GordonBurton1976a}, with a z-dependence
calculated using  two approximation at different Galactocentric distance R \cite{DickeyLockman1990,Cox1986}.
 The ionized component HII is calculated using a cylindrically symmetrical model \cite{Cordes1991}.\\
The Interstellar Radiation Field (ISRF) is calculated using emissivities based on stellar populations and dust emission \cite{Porter2005,Moskalenko2006}.\\
We consider a Galactic turbulent magnetic field with a mean value of the component perpendicular to the CR propagation given by

\begin{equation}\label{eq: Magn_field}
B=B_0  \exp{ \left( -\frac{ R-R_0 }{ R_{scale} } \right) } 
            \exp{ \left( -\frac{ |z| }{ z_{scale} } \right) },
\end{equation}

\noindent where $B_0=6~\mu G$, $R_{scale}=10~kpc$, $z_{scale}=2~kpc$ and $R_0=8.5~kpc$ the Sun Galactocentric distance.

\subsection{Diffusive Galactic spectrum}\label{Diffusive Galactic spectrum}

\begin{figure}
\begin{center}
\includegraphics [width=1\textwidth]{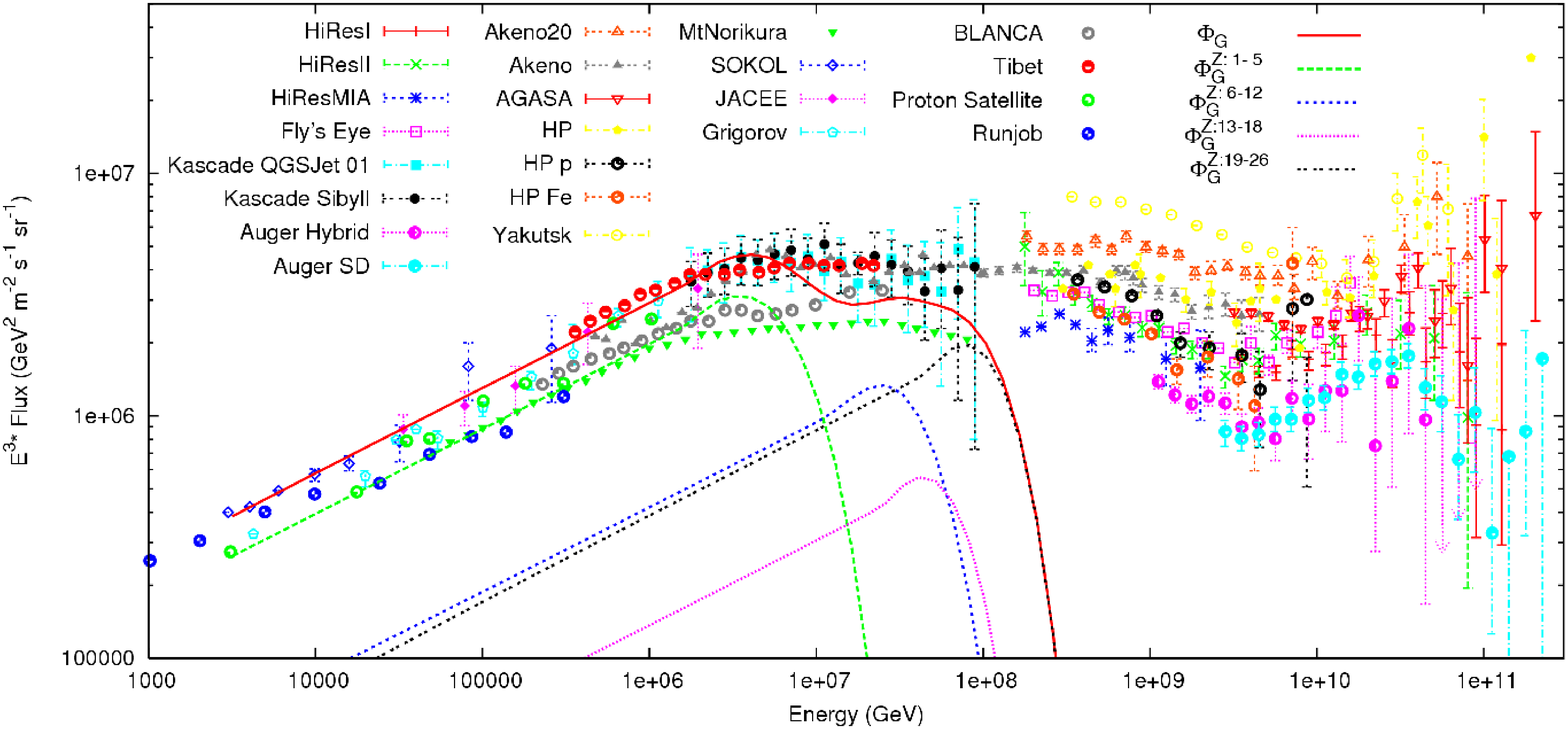}
\end{center}
\caption{Diffusive Galactic spectrum,  $\Phi_G$. The different Z-grouped nuclei components are shown.
}\label{GAL}
\end{figure}

The calculated diffusive Galactic spectrum is shown in Fig.\ref{GAL} superimposed to several experimental data results from 
the air shower experiments (HiRes \cite{HiRes2005b}, HiResMIA \cite{HiResMIA2001}, Fly's Eye \cite{FlysEye1994}, KASCADE \cite{Kascade2005},
Yakutsk \cite{Yakutsk2003}, Akeno \cite{Akeno1984,Akeno1992}, AGASA \cite{AGASA2003}, Haverah Park \cite{HP1991,HP2003},
Auger \cite{AugerFD_ICRC2007,AugerSD_ICRC2007}, BLANCA \cite{BLANCA2001}, Tibet \cite{Tibet2000,Tibet2003}, Mt. Norikura \cite{MtNorikura1997}) 
 as well as from direct measurements (SOKOL \cite{SOKOL1993}, JACEE \cite{JACEE1995,JACEE1998},
Grigorov \cite{Grigorov1999}, Proton Satellite \cite{ProtonSatellite1971}, Runjob \cite{RUNJOB2005}).

It can be seen that there is considerable dispersion over the whole energy interval, which highlights the inherent difficulty in CR spectral measurements over so many decades in energy. Furthermore, it can be seen that, below the first knee the dispersion is mainly limited to flux normalization, while the spectral indexes seem to be fairly consistent for different experiments. The situation worsens at higher energies until large discrepancies are apparent above $10^{17}$ eV. However, even at these high energies, a careful renormalization in energy \cite{Berezinsky2007ICRC} seems to be able to reconcile the several experimental data sets, giving a rather clear and consistent picture for the ankle. Unfortunately, normalization flux differences among spectra still remain between major experiments after the previous energy scale shifting, and they represent a concern to be experimentally addressed.  

The diffusive Galactic spectrum $\Phi_G$ is normalized to match KASCADE 
data at $\sim 3 \times 10^6$ GeV \cite{Kascade2005}. At this energy the differential flux value  
of the all particle energy spectrum and its uncertainty are the same for QGSJet 01 and Sibyll 2.1 based analysis of the KASCADE experiment.
While, with this renormalization, our spectrum agrees with data of various experiments  at lower energies 
(JACEE, SOKOL, Tibet, KASCADE, Haverah Park and Akeno),
beyond the knee the diffusive spectrum presents a strong deficit of
flux (Fig.\ref{GAL}).

\begin{figure}
\begin{center}
\includegraphics [width=1\textwidth]{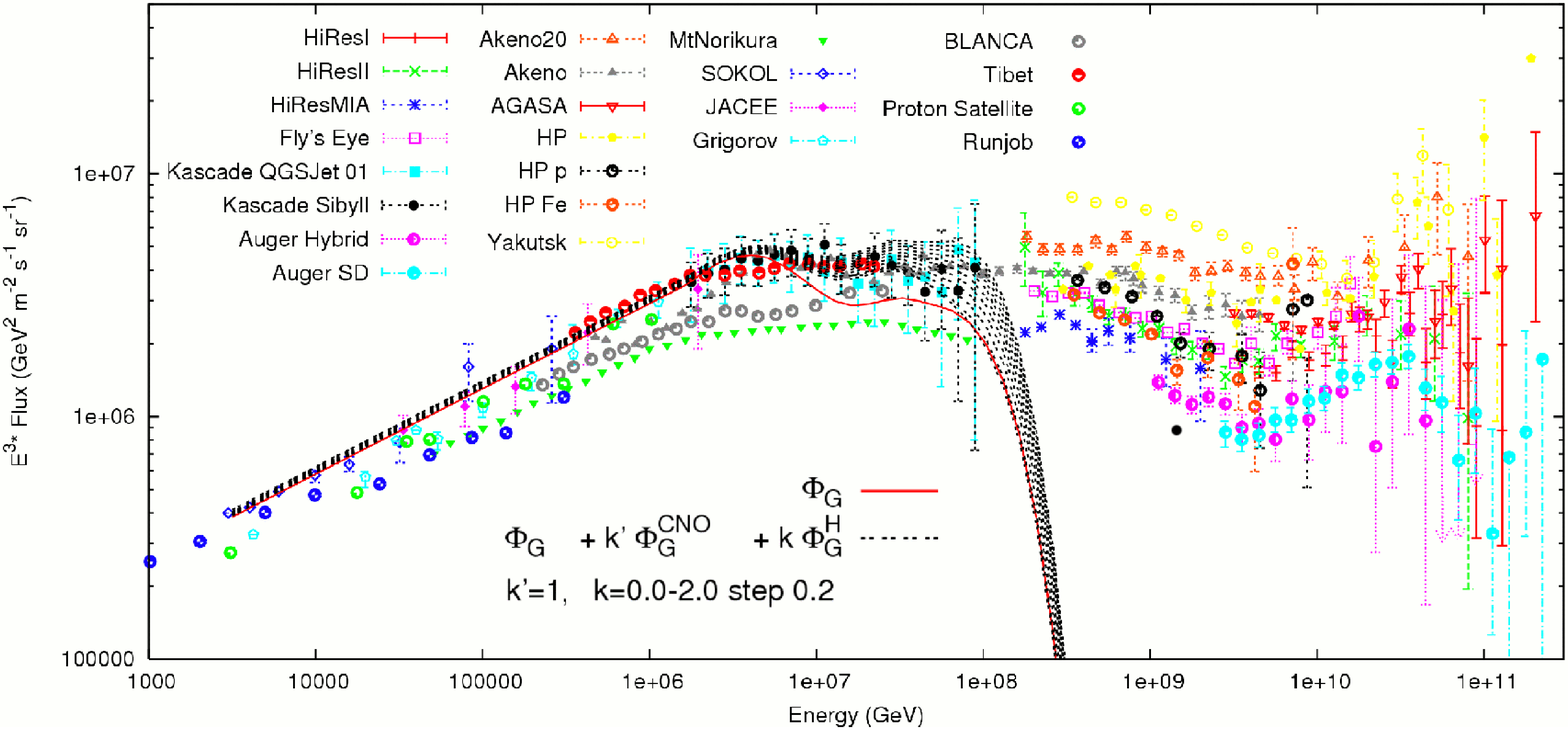}
\end{center}
\caption{Diffusive Galactic spectrum,  $\Phi_G$, with
renormalization of the CNO group ($Z:6-12$), $\Phi_G^{CNO}$, and of the
heavy component ($Z:19-26$), $\Phi_G^{H}$. Agreement with
experimental data is obtained for $k^\prime \approx 1$ and 
$k\approx0.8-1.2$.}\label{GALrenormC}
\end{figure}

Since at $E> 10^7$~GeV the composition is dominated by
intermediate ($Z:6-12$, i.e the CNO group) and heavier ($Z:19-26$) nuclei, we
renormalize these components ($\Phi_G^{CNO}$, $\Phi_G^H$) by a factor of $2$, which produces a
good agreement with the experimental data (see Fig.\ref{GALrenormC}).\\
The renormalization of the CNO group and of the heavy nuclei group 
by a factor of $2$ is acceptable since it corresponds to a renormalization of the injected abundances into the acceleration mechanism. 
The Galactic CR CNO nuclei are over abundant with respect to Iron by several orders of magnitude. 
Therefore, the Iron contribution to CNO CR flux resulting from spallation is small.
Consequently, to a good approximation, the observed Galactic abundances of Iron and the CNO group at the Solar circle can be varied independently by modifying their relative abundances at injection.
This amount of relative renormalization is well inside the present uncertainties regarding the detailed workings of the injection mechanism.

The renormalized diffusive Galactic spectrum reproduces well the
data up to $E\approx10^{8}~GeV$, beyond which the spectrum falls
steeply because of the end of the SNR acceleration efficiency.

\subsection{Extragalactic spectrum}\label{EG}

In order to study how the transition between the Galactic and
extragalactic components takes place, we compare the Galactic
spectrum originated in SNRs with two different possible scenarios for
the extragalactic component.

In the first model \cite{Berezinsky2006c}, a pure proton
extragalactic spectrum, accelerated by a homogeneous distribution of
cosmic sources, is considered. Local overdensities/deficits  of
UHECR sources affect the shape of the GZK modulation, but do not
affect the low energy region where the matching with the Galactic
spectrum occurs.
Within this model, we consider different cases of local overdensity/deficit of sources:
\begin{enumerate}
\label{DensityModel}
\item universal spectrum: $n/n_0=1$;
\item overdensity of sources: $n/n_0=2$,  $R=30~Mpc$;
\item overdensity of sources: $n/n_0=3$,  $R=30~Mpc$;
\item deficit of sources: $n/n_0=0$, $R=10~Mpc$;
\item deficit of sources: $n/n_0=0$, $R=30~Mpc$;
\end{enumerate}
where $n_0$ is the mean extragalactic source density and $n$ is the local overdensity/deficit in regions of size $R$.

In the second model \cite{Allard2007}, the extragalactic spectrum is
calculated for a mixed composition at injection typical of low
energy cosmic rays. The spectral index $\beta$ is determined fitting the high energy CR data.
Three different source evolution models in red shift are considered:
\begin{itemize}\label{EvolModel}

\item[a)] \emph{strong evolution model}: the injection rate is  proportional to $(1 + z)^4$
for $z<1$ and constant  for $1<z<6$, followed by a sharp cut-off, $\beta=2.1$;

\item[b)] \emph{SFR model}: the EGCR injection power is proportional to the star formation
rate which correspond to a redshift evolution  $(1 + z)^3$ for $z<1.3$ and a constant injection rate
for $1.3<z<6$ (with a sharp cut-off at $z = 6$), $\beta=2.2$;

\item[c)] \emph{uniform source distribution model}: no evolution, $\beta=2.3$.

\end{itemize}

In both cases, the various parameters of the models are tuned to fit
the available CR data at UHE and are, in that highest energy regime,
experimentally indistinguishable at present.
The spectra of both the models, used in the next sections, are renormalized to HiRes data.

\subsection{Combined spectrum: matching Galactic and extragalactic components}
\label{Combined spectrum}
In order to study how the transition between the Galactic and
extragalactic components takes place, we subtract the combined
theoretical (Galactic $\Phi_G$ plus extragalactic $\Phi_{EG}$) spectrum from the
available data. Two different approaches are used.

\begin{figure}
\begin{center}
\includegraphics [width=1\textwidth]{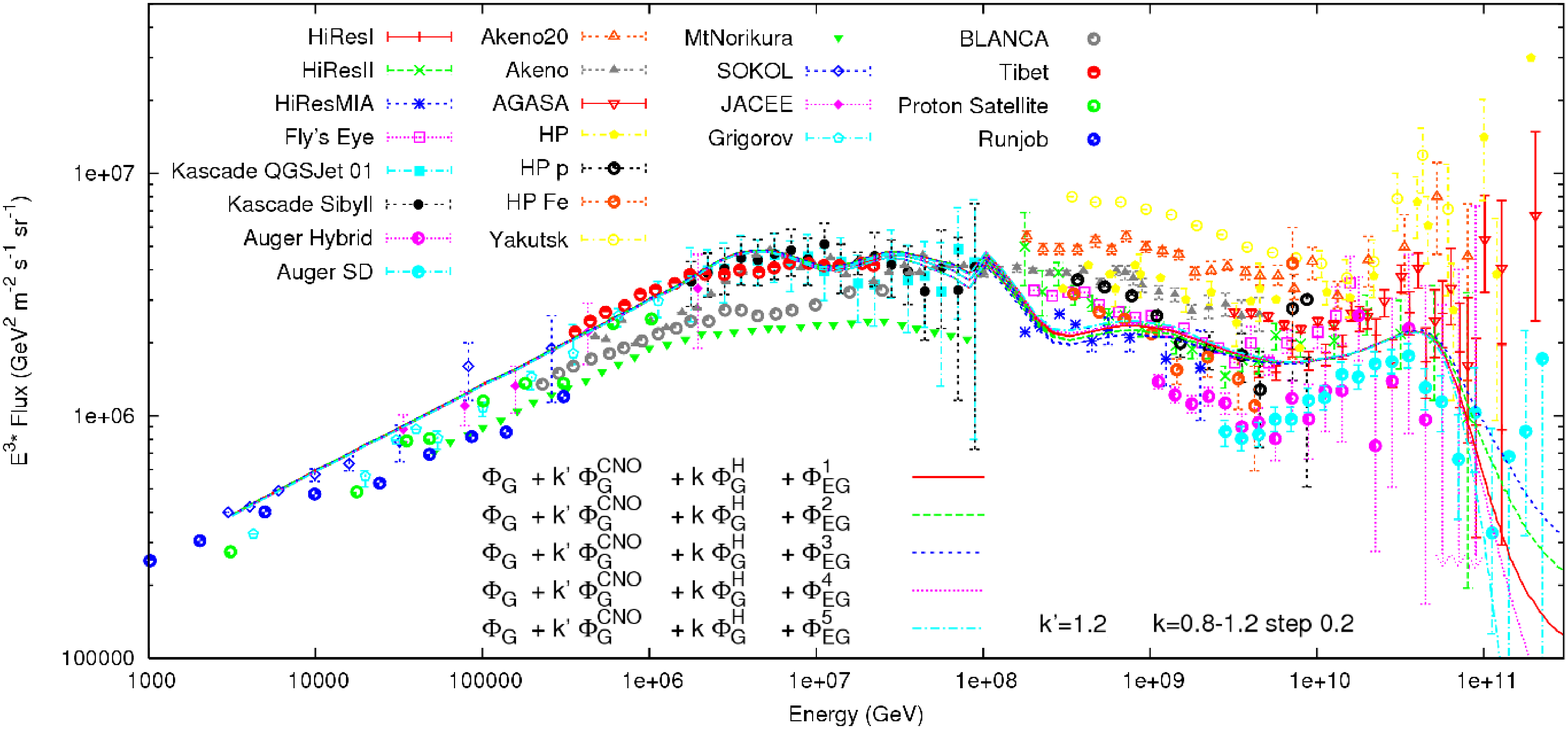}
\end{center}
\caption{Galactic and extragalactic spectrum matching for the {\it
proton\/} models for an EG lower energy limit of $10^8~GeV$. The sum
of the renormalized diffusive Galactic spectrum and of the
extragalactic spectrum ($\Phi_{EG}$) for different cases of local overdensity/deficit of sources (\S\ref{EG}) is
shown for different renormalizations of the heavy component
($\Phi_G^{H}$). The CNO group ($\Phi_G^{CNO}$) of the diffusive
Galactic spectrum ($\Phi_G$) has been renormalized by a factor
$2.2$.}\label{BeGR}
\end{figure}

\begin{figure}
\begin{center}
\includegraphics [width=1\textwidth]{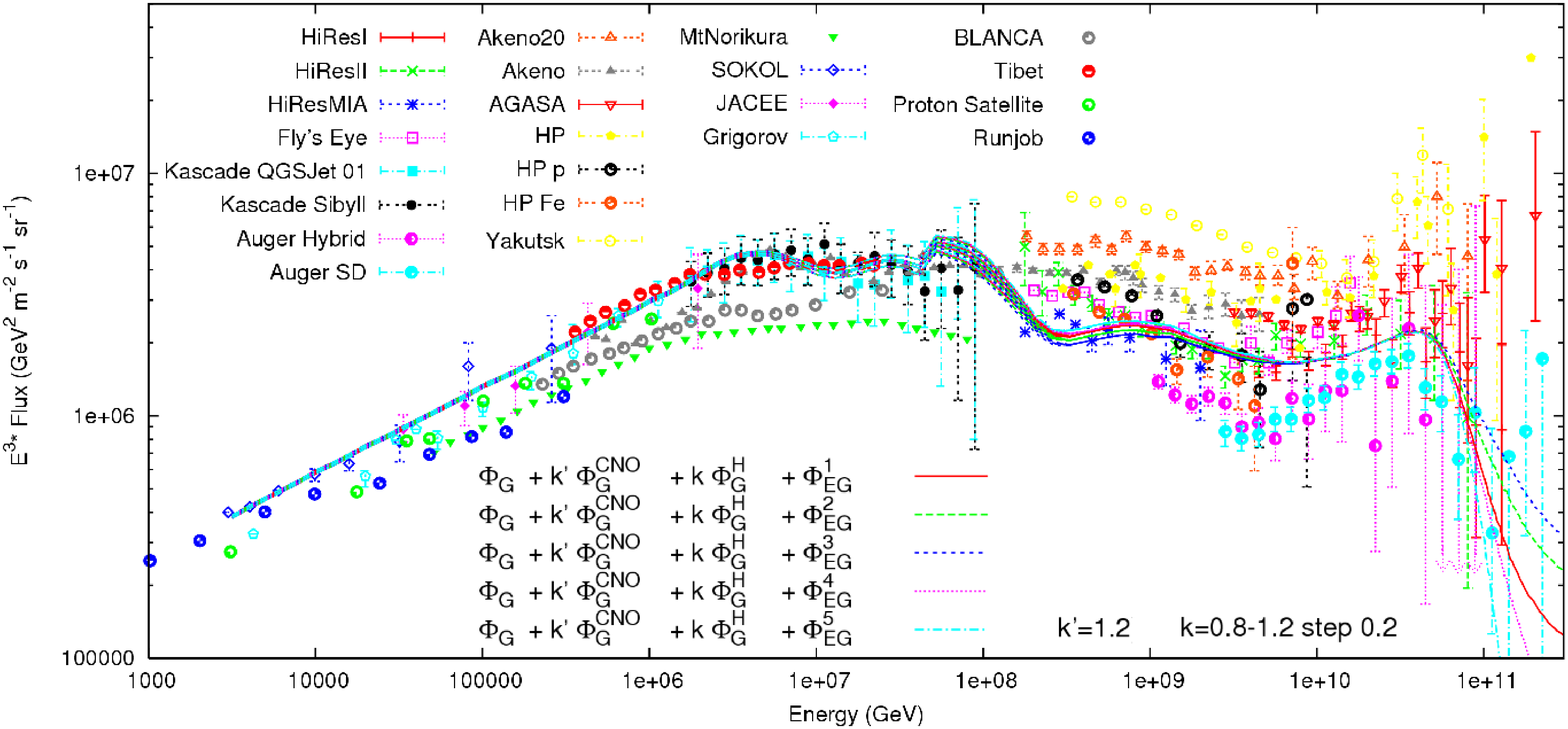}
\end{center}
\caption{Idem to Fig.\ref{BeGR} but for the EG {\it proton\/} model
with a lower energy limit of $5*10^7$~GeV.}\label{BeGRLL}
\end{figure}

\begin{figure}
\begin{center}
\includegraphics [width=1\textwidth]{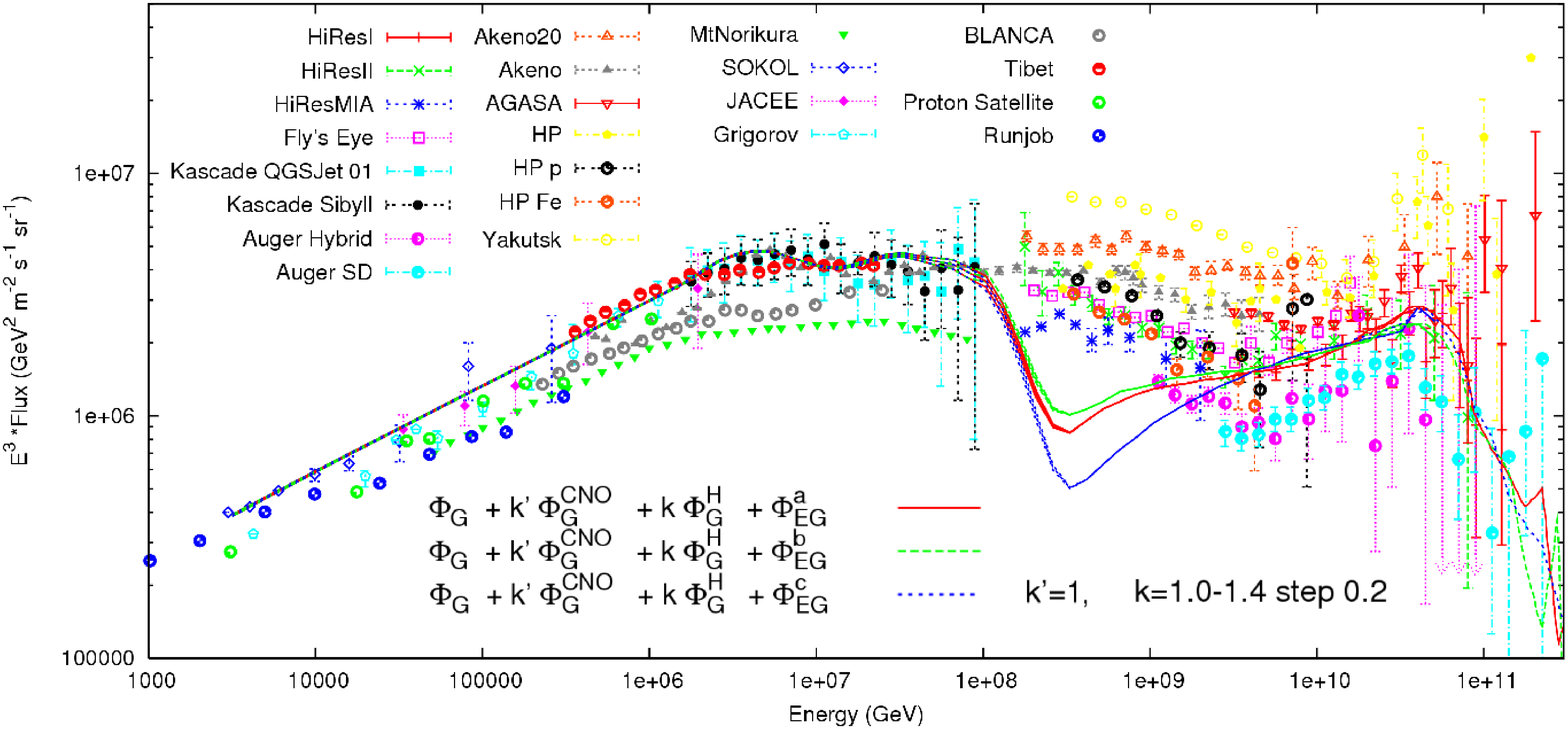}
\end{center}
\caption{Idem to Fig.\ref{BeGRLL} but for the Galactic and
extragalactic spectrum for the {\it mixed-composition\/} model. The
CNO group ($\Phi_G^{CNO}$) of the diffusive Galactic spectrum
($\Phi_G$) has been renormalized by a factor $2$. Three source
evolution models are considered (\S\ref{EG}): (a) strong, (b)
SFR and (c) uniform. }\label{AllGR}
\end{figure}

First, we try to match the experimental data by varying the
normalization of the heavy Galactic component $\Phi_G^H$, while keeping 
constant the previous renormalization of the CNO group $\Phi_G^{CNO}$ (\S\ref{Diffusive Galactic spectrum}).  
 The best reproduced spectrum for the two extragalactic models are shown in
Figs.\ref{BeGR}, \ref{BeGRLL} and \ref{AllGR}. In the case of the
{\it proton\/} model, a discontinuity appears when the two spectra
are added, regardless of the lower limit adopted for the
extragalactic component: $10^{8}$~GeV (Fig.\ref{BeGR}) or $5 \times
10^7$~GeV (Fig.\ref{BeGRLL}). The latter corresponds to cosmic
accelerators operating for the entire Hubble time.\\
For both, {\it proton\/} and {\it mixed-composition\/} models, there
is a flux deficit above $10^8$~GeV. The problem is much stronger for
the {\it mixed-composition\/} model where, regardless of the
luminosity evolution of EG CR sources, the total spectrum presents a
large deficit of flux between $10^8$ GeV and $\approx 3 \times
10^9$~GeV (Fig.\ref{AllGR}).\\

\begin{figure}
\begin{center}
\includegraphics [width=1\textwidth]{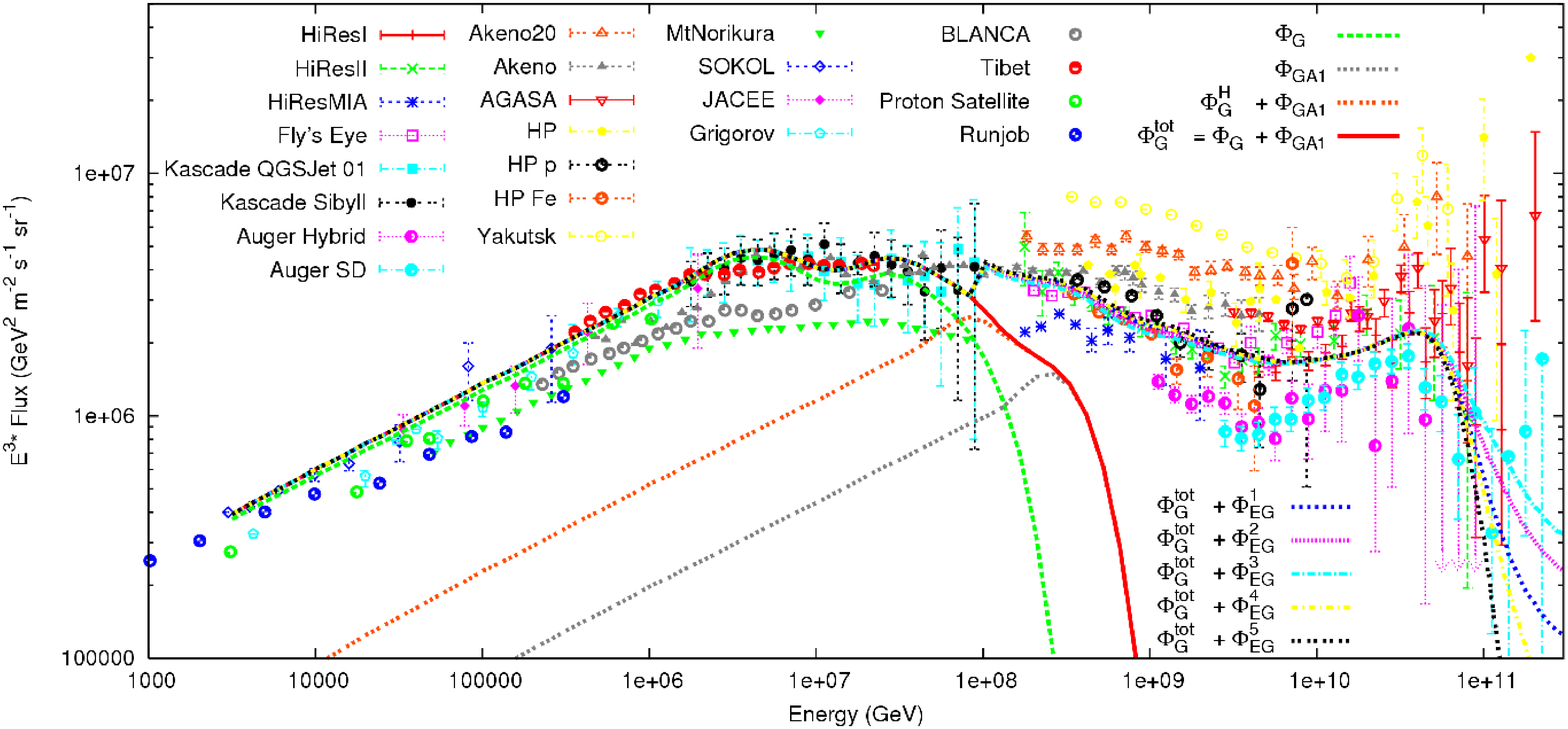}
\end{center}
\caption{Galactic and extragalactic spectrum for the {\it proton\/}
model for an EG lower limit of $10^8~GeV$: the additional Galactic  
component $\phi_{GA1}$ and the total  high energy Galactic component ($\phi_{G}^H +
\phi_{GA1}$) are also shown. }\label{BeGAc}
\end{figure}

\begin{figure}
\begin{center}
\includegraphics [width=1\textwidth]{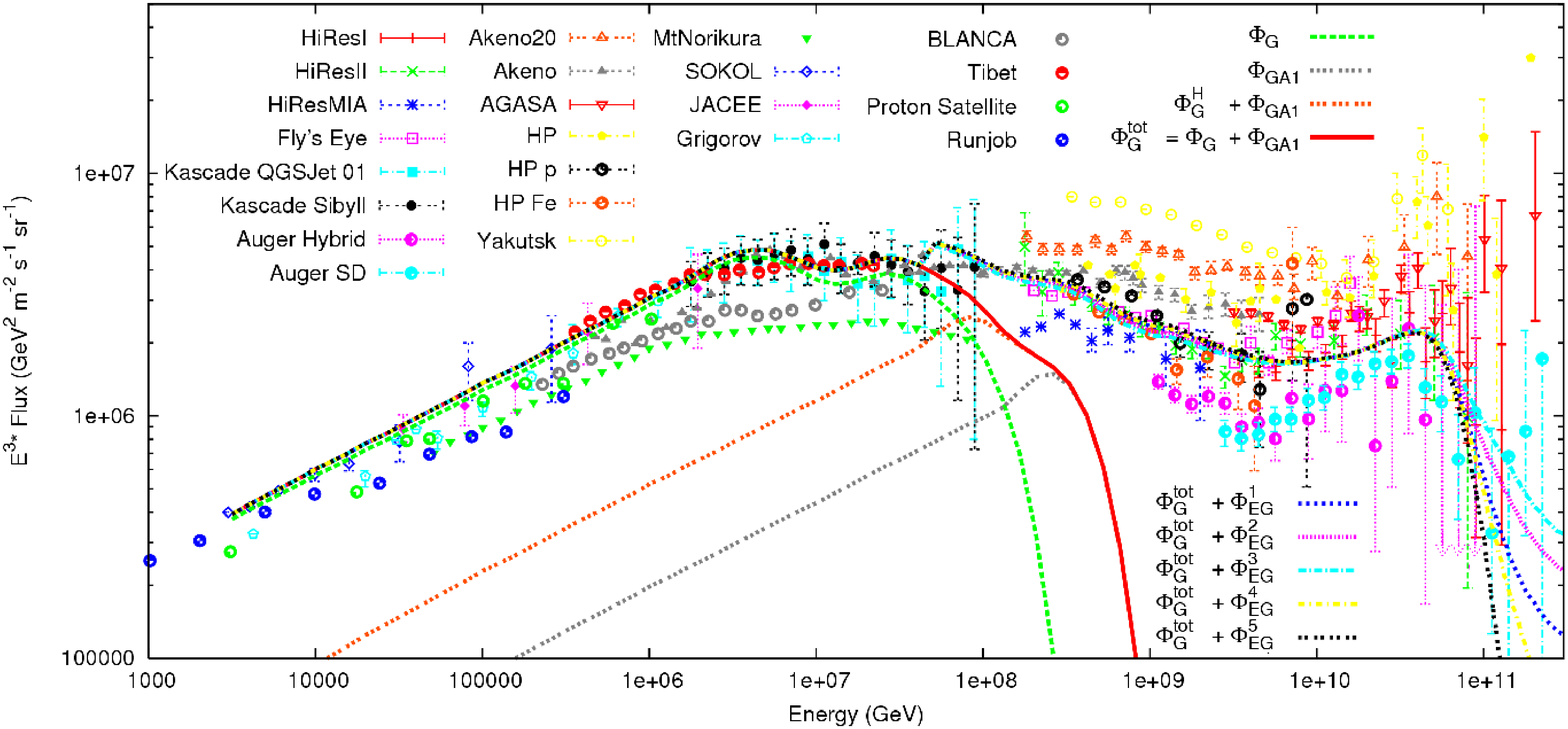}
\end{center}
\caption{Idem to Fig.\ref{BeGAc} but for the  EG  {\it proton\/} model with a lower energy limit of $5
\times 10^7~GeV$: the additional Galactic component $\phi_{GA1}$ and
the total high energy Galactic component ($\phi_{G}^H + \phi_{GA1}$) are also
shown.}\label{BeGALLc}
\end{figure}

\begin{figure}
\begin{center}
\includegraphics [width=1\textwidth]{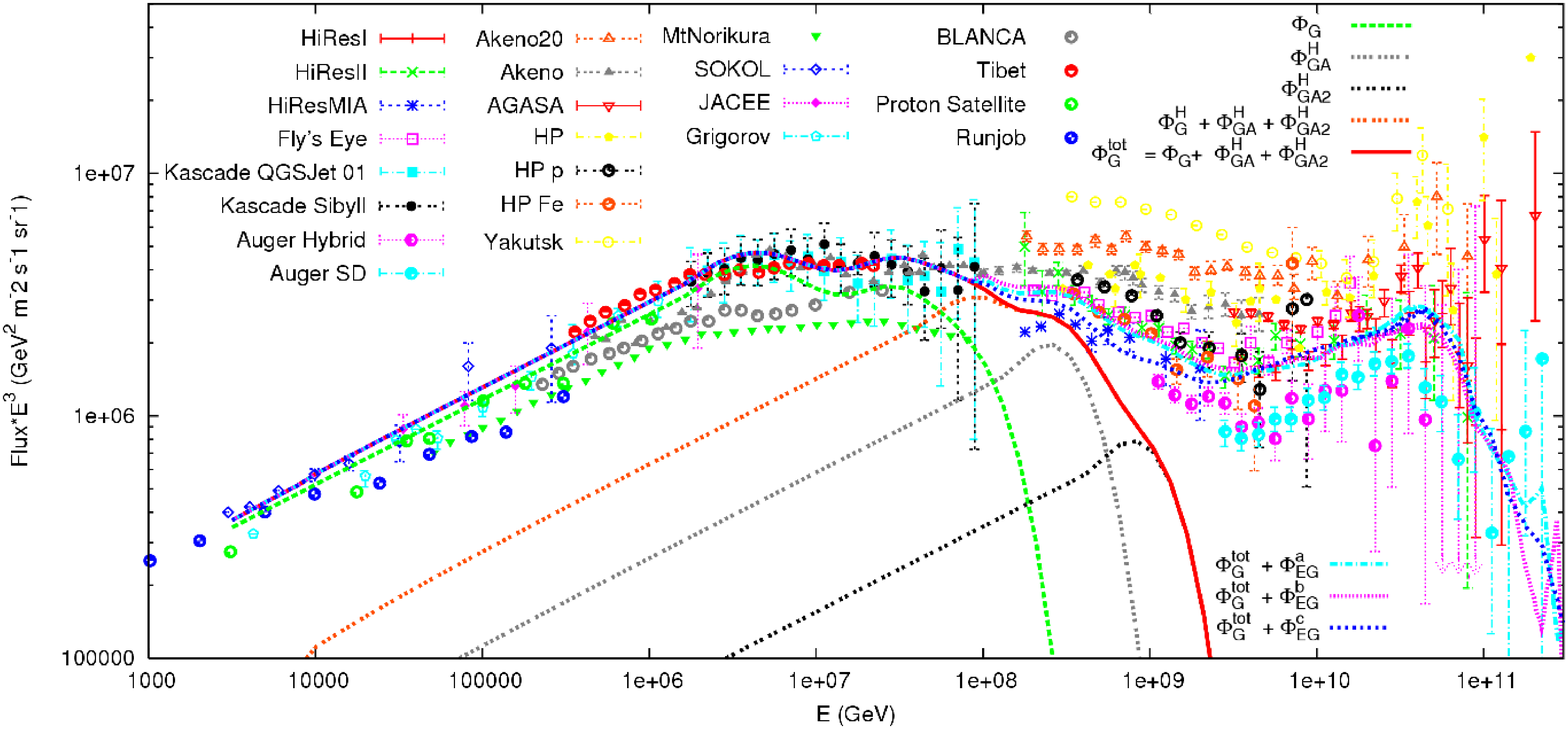}
\end{center}
\caption{Galactic and extragalactic spectrum for the {\it
mixed-composition\/} model: the two additional components
$\phi_{GA1}$, $\phi_{GA2}$ and the total high energy Galactic component
($\phi_{G}^H + \phi_{GA1} + \phi_{GA2}$) are shown.}
\label{AlGAc}
\end{figure}

In order to solve this flux deficit, the only way out seems to be
the introduction of an additional Galactic component. We estimate
this component by subtracting the sum of the diffusive Galactic and
extragalactic fluxes from a smooth fit to the world data. This
method confirms us the need of the renormalization of the CNO group
by a factor $\approx 2.2$ and $\approx 2$ for the {\it proton\/} and
{\it mixed-composition\/} models, respectively. In the case of the
{\it proton\/} models (Figs.\ref{BeGAc}, \ref{BeGALLc}), the
observed deficit can be resolved with the introduction of an additional cosmic 
ray flux component (the first additional Galactic component, GA1) $\Phi_{GA1}$.
For the {\it mixed-composition\/} models this is not enough and one more additional 
cosmic ray flux component must be introduced 
(the second additional Galactic component, GA2) $\Phi_{GA2}$ (Fig.\ref{AlGAc}). The
additional component $\Phi_{GA1}$ common to both families of models is obtained
with a shift in energy of a factor $\approx 3$ of the diffusive
Galactic heavy component, renormalized by a factor $\approx 0.8$ in
the case of the {\it mixed-composition\/} models and $\lesssim0.6$
for the {\it proton\/} models respectively. The second additional
component $\Phi_{GA2}$ is obtained in an analogous way but with an energy-shift
factor of $\approx 10$ and a renormalization  by a factor $\approx0.2$. The corresponding spectra are shown in Figs.\ref{BeGAc},
\ref{BeGALLc} and \ref{AlGAc}.

\subsection{Discussion on the spectrum}

We have analyzed the matching conditions of the Galactic and
extragalactic components of cosmic rays along the second knee and
the ankle.\\
It seems clear that an acceptable matching of the Galactic and
extragalactic fluxes can only be achieved if the Galaxy has
additional accelerators, besides the fiducial SNRs assumed here, 
operating in the interstellar medium.

 In the particular case of the {\it proton\/} model, only one additional component is required. 
This could well represent the contribution from compact and highly magnetized
SNRs, like those occurring in the central, high density regions of
the Galactic bulge, inside the dense cores of molecular clouds or those expanding
into the circumstellar winds of their progenitors. Actually, 
this component does not need to originate in a particular type of source
in itself, but could be the result of a non-homogeneous SNR population 
drawn from a spectrum of progenitor masses and evolving in different 
environments corresponding to the various gas phases that fill the interstellar 
medium \cite{Stanev1993}.  
From the point of view of CR luminosity this should not be a problem,
since the Fe component and the additional high energy component 
amount to $\sim 9$\% and $\lesssim 1$\%, respectively, of the total Galactic
diffusive spectrum.

In the specific case of the {\it proton \/} model, it is apparent that a perfectly 
smooth match between the Galactic and the extragalactic spectra 
is not possible. Some discontinuity or wiggle seems unavoidable in the 
energy spectrum between $\sim 3 \times 10^{16}$ and $\sim 10^{17}$ eV 
(see figures \ref{BeGR}, \ref{BeGRLL}, \ref{BeGAc} and \ref{BeGALLc}). 
This is due to the fact that the extragalactic spectrum should have a 
rather abrupt low energy cut-off due to the finite distance to the nearest 
extragalactic sources and their limited age. Of course, whether such 
spectral feature is actually observable is strongly dependent 
in practice on the magnitude of the experimental errors in the 
determination of the primary energy from shower measurements 
and on the available statistics.\\

The matching of the {\it mixed-composition\/} model, on the other hand, 
has wider astrophysical implications for the Galaxy.
The Galactic CR production has to be extended up to at least the
middle of the dip and this requires, besides the previous additional
component, another high energy Galactic component. The origin of
these cosmic rays pushes even further the acceleration requirements
imposed on the Galaxy. It is likely that a different population of 
Galactic accelerators must be invoked. Viable candidates could include,
for example, rapidly spinning inductors, like pulsars or magnetars, or even
very high energy episodic events like Galactic gamma ray bursts \cite{GMTancoGRB}. 
If this were the case, it is very likely that photon emission at TeV 
energies should uncover the sources. Changes in propagation regime 
inside the Galaxy at these high energies should manifest as an  
anisotropic component embedded in the isotropic incoming extragalactic 
background. Again, considerable statistics might be necessary to make
such effect observable.

The energy requirements involved in the production of the second additional component
are rather modest: $\sim 10^{-7}$ of the SNR CR component, or $\sim 10^{34}$ 
erg/sec pumped into particles between $\sim 10^{17}$ and $\sim 10^{18}$ eV.
Therefore, few, or even a single source, could be responsible for this component.
This carries attached the potential problem of undesirable fine-tuning because of
the requirement that, at this precise moment in time, the CR luminosity of these
few (or this single) sources is such that it allows for a smooth spectral matching
along the ankle.\\

It is important to note that the region between some few times $10^{17}$ eV 
and approximately $10^{19}$ eV is a transition region from the point 
of view of propagation \cite{GMTanco2006,GMTanco2007}. In fact, 
Larmor radius (in pc) of a CR nucleus of charge $Z$ can be conveniently 
parametrized as: 

\begin{equation} \label{LarmorRadius}
r_{L,pc} \sim \frac{10}{Z} \times \left( \frac{E_{EeV}}{B_{\mu G}} \right)
\end{equation}

\noindent where $E_{EeV}$ is its energy and $B_{\mu G}$ is some 
appropriate average of the Galactic magnetic field inside the propagation 
region. Since the transversal dimensions of the Galactic disk are on the 
order of a few times $10^2$ pc, we see that the diffusion approximation for 
protons should start to be broken somewhere between the second knee
and $1$ EeV. The same should happen to heavier nuclei at progressively 
higher total energies: $\sim 3$ EeV for the CNO group and $\sim 10$ EeV
for Fe. The corresponding transition for each nuclei should be gradual, with 
the propagation eventually becoming ballistic at higher energies. This 
transition results in a complicated picture in which the end of the Galactic 
confinement spans almost $1.5$ decade in energy, depending on the particular 
nuclei considered. Our interpretation of the additional Galactic components 
(but not their necessity) is affected to some extent by the implicit assumption 
that the particles propagate diffusively. This assumption is increasingly 
questionable for protons as we approach $1$ EeV, but should be valid for 
heavier nuclei in the energy range considered here. We are currently working 
on a detailed analysis of these effects which will be presented elsewhere.

\section{Composition}

The different extragalactic models are able to produce total spectra that are indistinguishable within the current experimental resolution. 
The composition of UHECR is essential to understand the transition between Galactic and extragalactic cosmic rays and to discriminate the different astrophysical scenarios.
 
Since UHECR experiments do not directly measure the composition, we have 
to infer it from parameters characterizing the shower development profile at ground. 
One of the most reliable parameters is  $X_{max}$, the atmospheric depth of the 
maximum longitudinal development of a shower \cite{Supanitsky2008a}. The variation of  $\langle X_{max}\rangle$ with energy gives 
information about the change in composition of the CR flux.

Different hadronic interaction models (HIMs) can be used to interpret the $X_{max}$ 
dependence on energy and on primary composition, and this is the main uncertainty 
associated with this parameter. In this section we will use different parameterizations of 
$\langle X_{max}\rangle$ deduced from the hadronic interaction models currently in use,
in order to infer the composition energy profile for the different cases of Galactic-extragalactic combined spectrum described in section \S\ref{Combined spectrum}.

\subsection{Hadronic interaction models}
\label{Hadronic interaction models}

The composition predictions depend on the HIM used. For
the sake of completeness, we take into account in what follows all the HIMs 
currently used in the literature:
\begin{itemize}
\item EPOS 1.6
\item QGSJet 01
\item QGSJetII v02
\item QGSJetII v03
\item Sibyll 2.1
\end{itemize}
For each hadronic interaction model, we calculate a parameterization of 
$\langle X_{max}\rangle$ as a function of the hadronic interaction model 
and of the primary energy and composition:
\begin{eqnarray} \label{eq:X_max}
\langle X_{max}\rangle (E, A, i) & = &
  \left[ a_i\left(Log\frac{E}{A\epsilon}\right)^2+ b_i Log\frac{E}{A\epsilon}+c_i\right](1+\alpha_i A)\\
& & + p_i(1+\beta_i A),\nonumber
\end{eqnarray}
\noindent where $E$ is the primary energy, $A$ is the atomic mass of the primary 
CR, $i$ is the hadronic interaction model,  $\epsilon\sim81~MeV$ is the critical 
energy in air  and $a_i,~b_i,~c_i,~p_i,~ \alpha_i,~\beta_i$ are coefficients determined 
by fitting equation \ref{eq:X_max} to shower simulations (see Table \ref{HadIntModels_Par}). 

\begin{table}[]
\vspace{0.4 cm}
\centering
\caption{\label{HadIntModels_Par} {Coefficients of the parameterization of  $\langle X_{max}\rangle$ as a function of primary energy and composition (eq. \ref{eq:X_max}) for different hadronic interaction models.}}
\vspace{0.2 cm}
\begin{tabular}{|c|r|r|r|r|r|r|}
\hline
\textbf{Model}& \textbf{$a$} & \textbf{$b$} & \textbf{$c$} & \textbf{$p$} & \textbf{$\alpha$} & \textbf{$\beta$}\\
\hline
QGSJetII-v03 & $-0.033$ & $-0.010$ & $58.313$ & $184.315$ & $9.60 \times10^{-4}$ & $-2.41 \times10^{-3}$\\
QGSJetII-v02 & $0.184$ & $-5.930$ & $115.111$ & $-15.895$ & $2.00 \times10^{-3}$ & $6.50\times 10^{-2}$ \\
QGSJet01 & $0.074$ & $-3.695$ & $103.804$ & $-19.123$ & $6.85 \times10^{-4}$ & $2.14 \times10^{-2}$ \\
EPOS 1.6 & $-0.011$ & $1.831$ & $29.212$ & $265.847 $ & $-1.47 \times10^{-4}$ & $4.69 \times10^{-4}$ \\
Sibyll 2.1 & $0.300$ & $-8.422$ & $134.960$ & $-72.051$ & $3.38 \times10^{-4}$ & $2.62 \times10^{-3}$ \\

\hline
\end{tabular}
\vspace{0.4 cm}

\end{table}

\begin{figure}
\begin{center}
\includegraphics [width=1\textwidth]{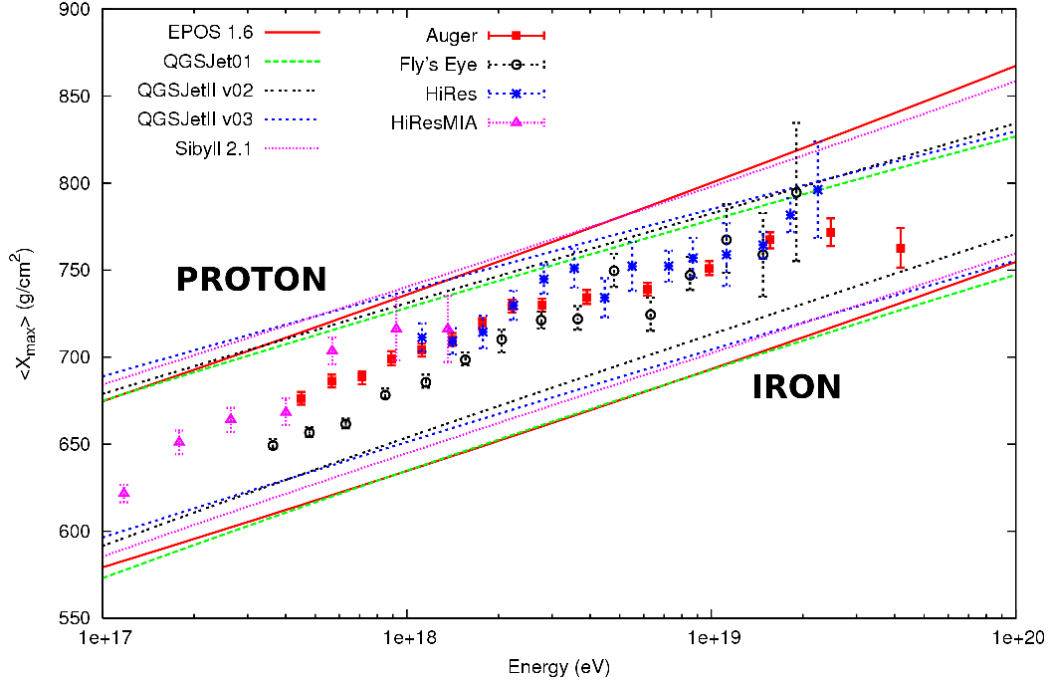}
\end{center}
\caption{$\langle X_{max}\rangle$ parameterization as a function of primary energy and composition. For each HIM, the $\langle X_{max}\rangle$ dependence on energy for proton and Iron primary is shown. Experimental data from HiRes, HiResMIA, Fly's Eye and Auger are also shown.}\label{X_max_models}
\end{figure}

Parameterization \ref{eq:X_max} is a good approximation in the energy range $[10^{17}~eV,~10^{20}~eV]$. Our parameterization for p and Fe, for several HIMs, is shown in Fig.\ref{X_max_models} together with the experimental results of HiRes \cite{HiRes2005a}, HiResMIA \cite{HiResMIA2001}, Fly's Eye \cite{FlysEye1993} and Auger \cite{Auger_Xmax_ICRC2007}.

This parameterization allows us to compute the average $X_{max}$ for showers 
generated by any nucleus of interest and for all observed primary energies above 
the second knee.

With this parameterization, in the following sections we reproduce $\langle X_{max} \rangle$ 
energy profiles for different composition scenarios in the context of the various cases of the combined 
Galactic-extragalactic spectrum.

\subsection{Galactic-extragalactic combined spectra: $X_{max}$ energy profiles}\label{XmaxProfiles}

Heretofore, we have analyzed from the point of view of the energy spectrum, different 
scenarios for the combination of the Galactic flux with alternative extragalactic models. 
For each of these scenarios, and under different assumptions for the composition of 
individual flux components, we estimate $X_{max}$ along the transition 
region.

As seen in \S\ref{Combined spectrum}, an acceptable matching of the Galactic and
extragalactic fluxes can only be achieved if the Galaxy has
additional accelerators besides the regular SNRs assumed in \S\ref{DiffModel}. 
This can be achieved by including either one (pure {\it proton\/} EG spectrum) or two 
({\it mixed-composition\/} EG spectrum) additional Galactic components. 

In order to infer $X_{max}$ for the combined flux, we have to do some 
assumptions on the composition of the additional Galactic components.

The first additional Galactic component (GA1), which is needed in both of the EG scenarios 
previously considered, is probably contributed by compact, highly magnetized 
SNRs. Consequently this component is likely to be dominated by heavy elements,
say, Iron, at the highest energies.

The second additional Galactic component (GA2), on the other hand, might be 
associated with a minor population of still more powerful SNR accelerators or
with a completely different population of particle accelerators, e.g., rapidly spinning inductors, 
like pulsars or magnetars. In the first case, GA2 would be Fe dominated, while in the latter case, 
GA2 could be proton dominated. Consequently, we consider two different possibilities
for GA2: pure proton and pure Fe.

\begin{figure}
\begin{center}
\includegraphics [width=1\textwidth]{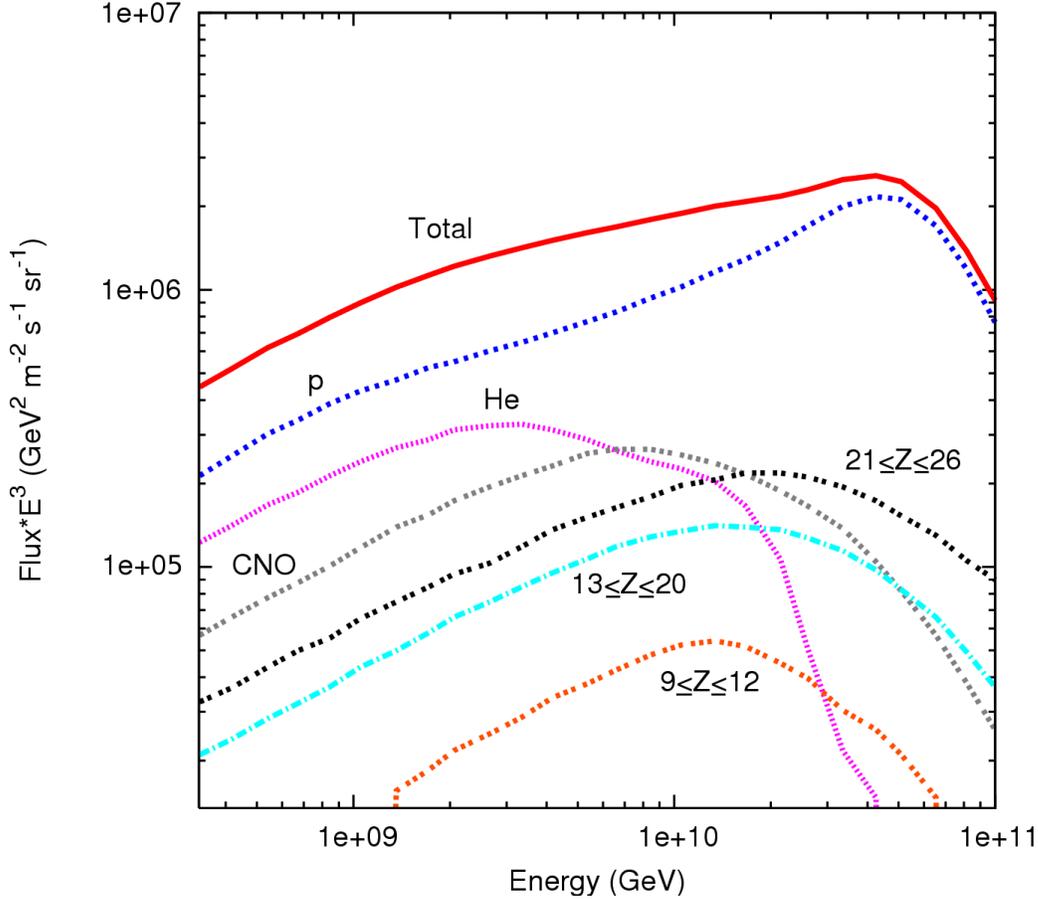}
\end{center}
\caption{{\it Mixed-composition\/} EG scenario of Allard and co-workers. Contribution 
to the extragalactic flux of different nuclei in the ``uniform source distribution model". (Adapted from: \cite{Allard2005b}).}\label{Allard_composition}
\end{figure}

We calculate $X_{max}$ as a function of energy in the energy range $[10^{17}~eV,10^{20}~eV]$ for all the possible combination of Galactic and extragalactic spectra previously discussed in \S\ref{Combined spectrum} from the point of view of the shape of the energy spectrum\begin{footnote}{For the diffusive Galactic spectrum from SNRs we kept the renormalization of the CNO group by a factor $\sim2$ determined in the previous analysis (\S\ref{Combined spectrum})}\end{footnote}:

\begin{itemize}

\item pure {\it proton\/} EG model: we consider the combination of the calculated Galactic spectrum from SNRs with the EG ``universal" (see \S\ref{EG}) 
spectrum in the two cases of lower EG energy limit ($5\times10^{16}~eV and \sim 10^{17}~eV$); the composition of the Galactic additional component is assumed to be heavy (Iron);

\item mixed EG composition model:  we consider the combination of the calculated Galactic spectrum from SNRs with the {\it mixed\/} 
 EG spectrum in its ``uniform source distribution model"  variant (see \S\ref{EG}); 
 we also assume that GA1 is composed by pure Fe, while we analyze two opposite scenarios for GA2: 
 pure proton and pure Fe. The EG composition is taken from Fig.\ref{Allard_composition} \cite{Allard2005b}.

\end{itemize}

For all these scenarios with different component spectra and composition assumptions, 
we calculate $\langle X_{max}\rangle$ values in the energy range 
$[10^{17}~eV,10^{20}~eV]$ using the parameterization given in \S\ref{Hadronic interaction models} for the various HIMs in current use.

\begin{figure}
\begin{center}
\includegraphics [width=1\textwidth]{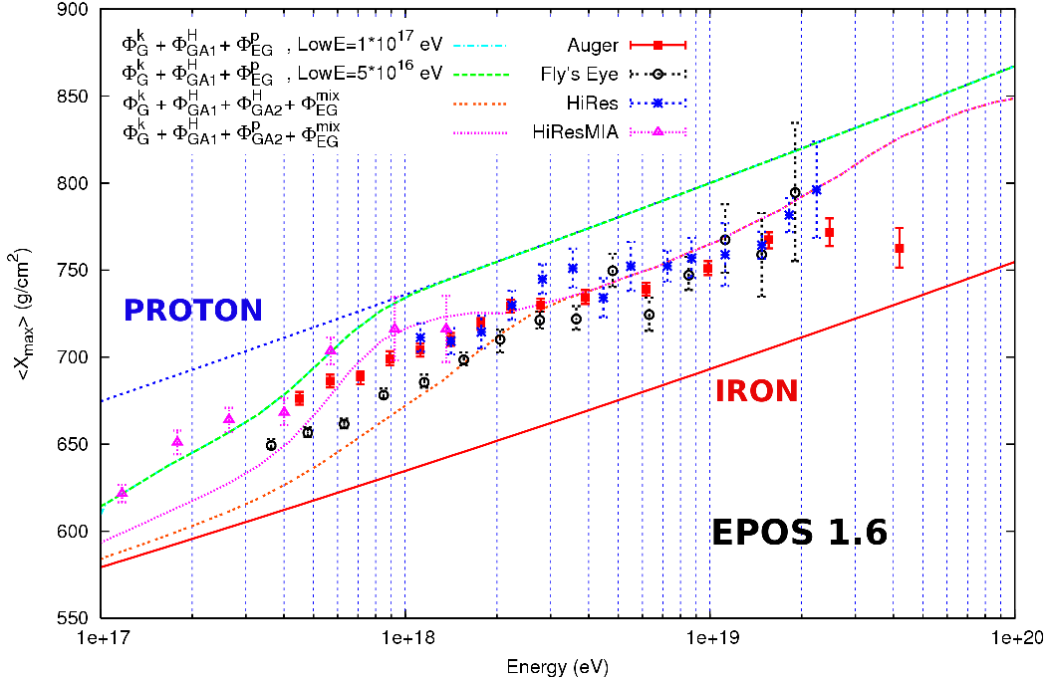}
\end{center}
\caption{$X_{max}$, EPOS 1.6: the $X_{max}$ profiles calculated for all the possible combined 
spectra are shown superimposed onto the experimental data. 
The curves represent: (i) the combined spectra for an EG {\it proton\/} model with lower energy limit  
 $\sim 10^{17}~eV$ (cyan blue) and $5\times 10^{17}~eV$ (green) respectively; (ii) the combined spectra for an EG {\it mixed-composition\/} model with pure proton (purple)  and pure Fe (brown-orange) GA2. $\langle X_{max}\rangle$ values are calculated using the EPOS 1.6 HIM.}\label{EPOS_X_max_Allard_Bere}
\end{figure}

\begin{figure}
\begin{center}
\includegraphics [width=1\textwidth]{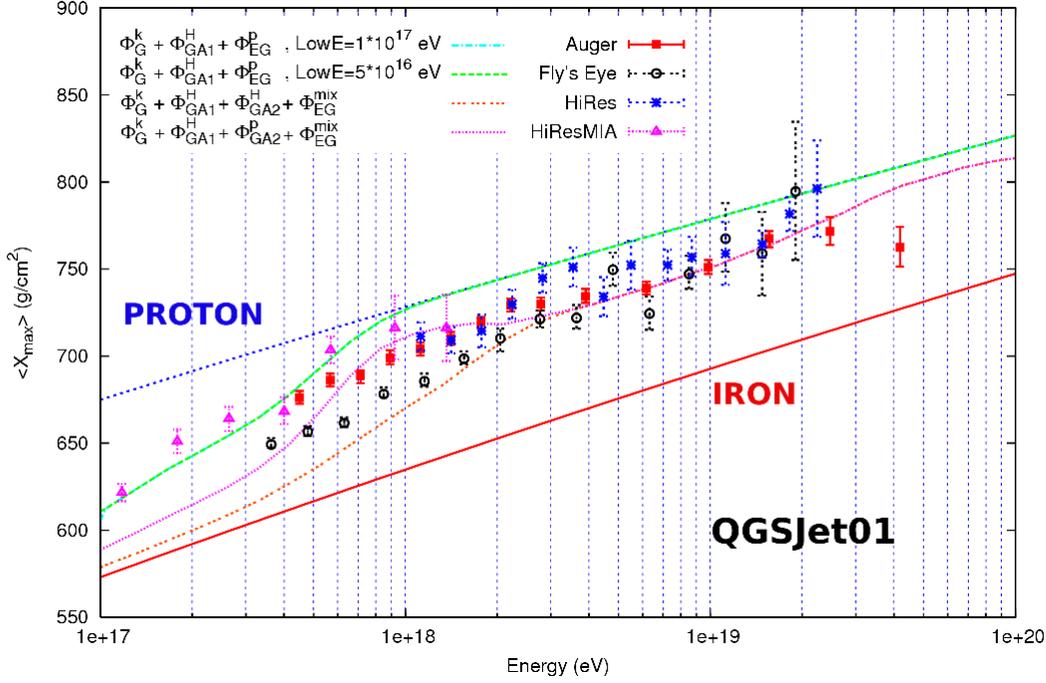}
\end{center}
\caption{$X_{max}$, QGSJet01: idem to Fig.\ref{EPOS_X_max_Allard_Bere} 
but using the  QGSJet01 HIM.}\label{QGSJet01_X_max_Allard_Bere}
\end{figure}

\begin{figure}
\begin{center}
\includegraphics [width=1\textwidth]{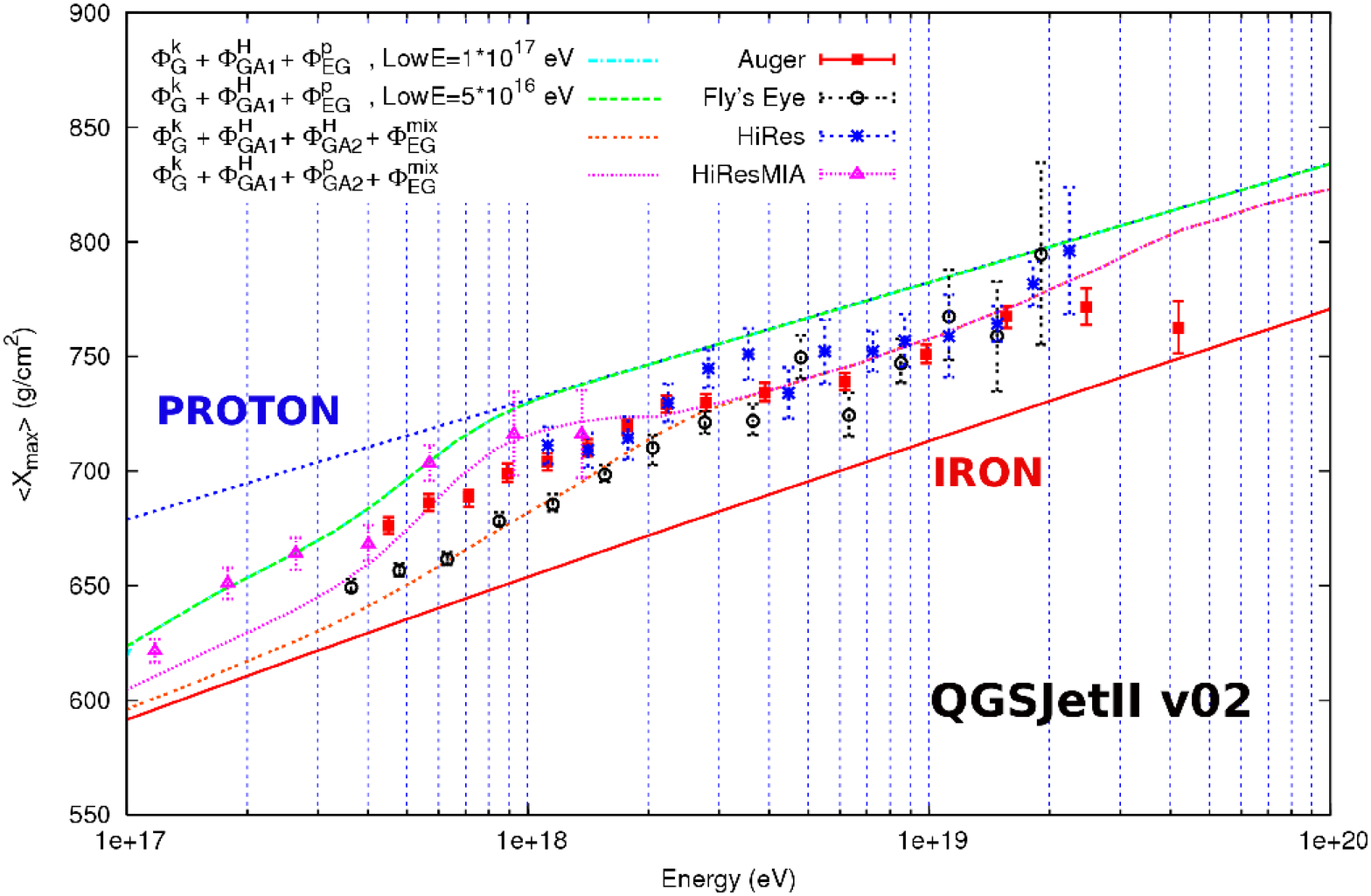}
\end{center}
\caption{$X_{max}$, QGSJetII v02: idem to Fig.\ref{EPOS_X_max_Allard_Bere}
but using the QGSJetII v02 HIM.}\label{QGSJetII_v02_X_max_Allard_Bere}
\end{figure}

\begin{figure}
\begin{center}
\includegraphics [width=1\textwidth]{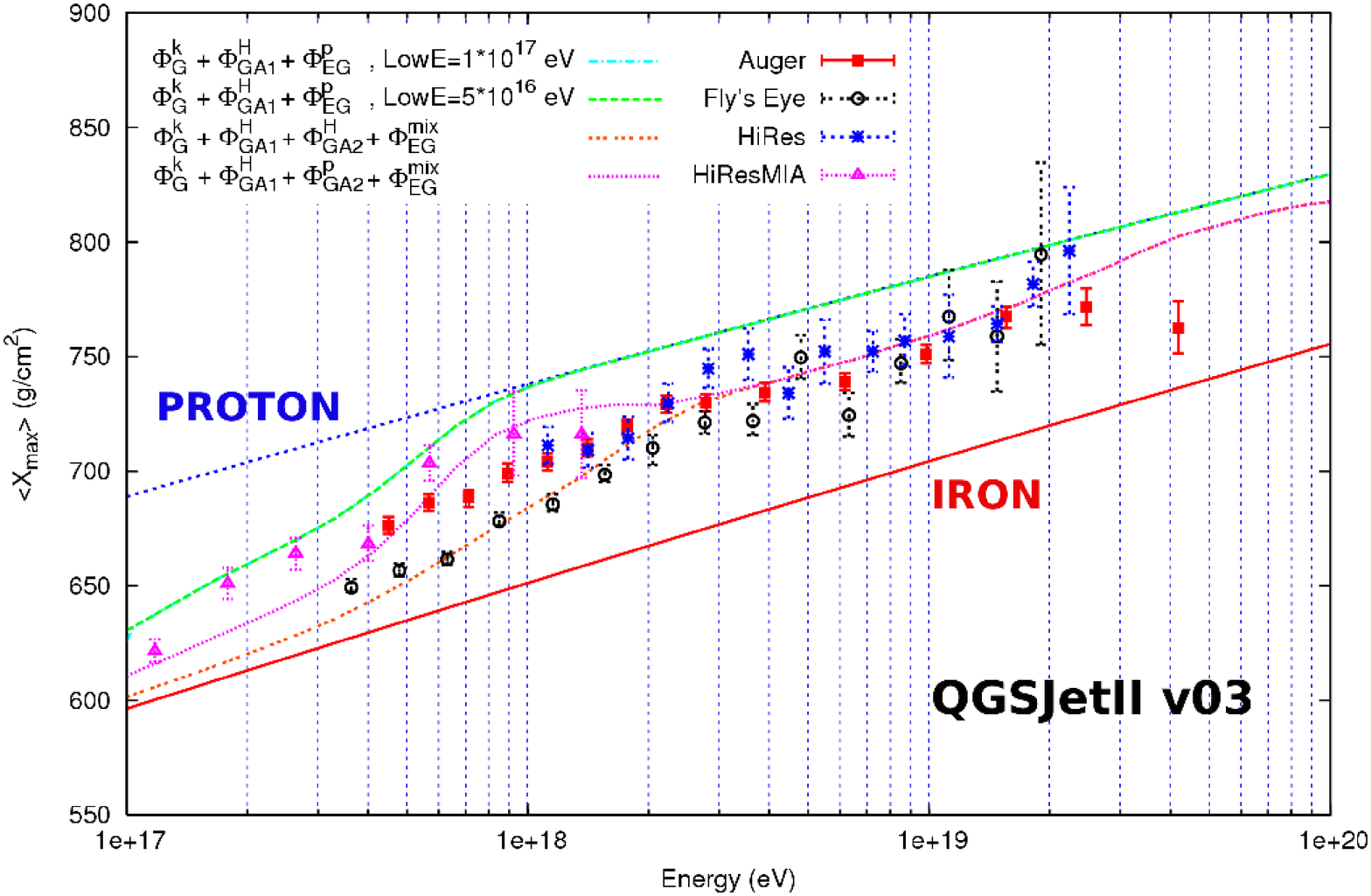}
\end{center}
\caption{$X_{max}$, QGSJetII v03: idem to Fig.\ref{EPOS_X_max_Allard_Bere}
but using the QGSJetII v03 HIM.}\label{QGSJetII_v03_X_max_Allard_Bere}
\end{figure}

\begin{figure}
\begin{center}
\includegraphics [width=1\textwidth]{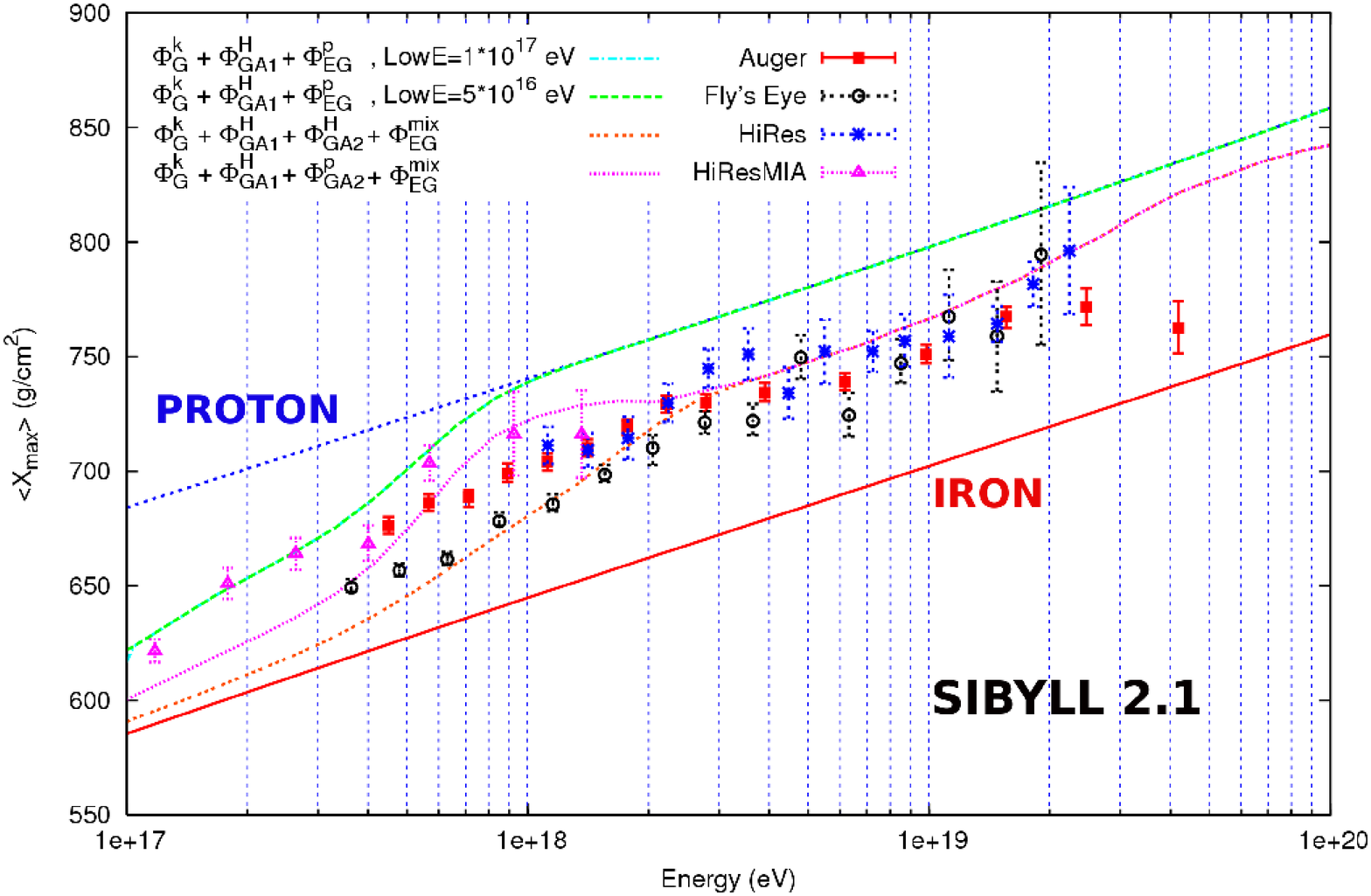}
\end{center}
\caption{$X_{max}$, Sibyll 2.1: idem to Fig.\ref{EPOS_X_max_Allard_Bere}
but using the Sibyll 2.1 HIM.}\label{Sibyll_X_max_Allard_Bere}
\end{figure}

The results are shown for each HIM in Figs.\ref{EPOS_X_max_Allard_Bere}, \ref{QGSJet01_X_max_Allard_Bere}, \ref{QGSJetII_v02_X_max_Allard_Bere},  
 \ref{QGSJetII_v03_X_max_Allard_Bere} and \ref{Sibyll_X_max_Allard_Bere}, 
 together with the experimental results of HiRes \cite{HiRes2005a}, HiResMIA \cite{HiResMIA2001},
Fly's Eye \cite{FlysEye1993} and Auger \cite{Auger_Xmax_ICRC2007}.

First of all we notice that, as expected, for all the HIMs the theoretical $X_{max}$ values in the case of EG {\it proton\/} model 
 are indistinguishable for the two different EG lower energy limits (green and cyan blue curves are superimposed on the figures). 
 On the other hand, the model is very much compatible with HiRes-MIA data below $10^{18}$ eV while, at higher energies, 
 $\langle X_{max} \rangle$  is too large when compared to any of the available data sets.

As expected, much more variability in the calculated elongation rates is observed for the lower energy 
 regime in the case of EG {\it mixed-composition\/} models, depending on the assumptions made about the composition of GA2 (purple and brown-orange curves). However, at energies larger than $\sim 3 \times 10^{18}$ eV, all the solutions converge to the same profile since the effects of GA2 become progressively negligible on the combined composition profile. All these results are qualitatively independent of the assumed HIM. Nevertheless, even if all the HIMs give basically the same trend at all energies, actually what data 
set is quantitatively compatible with the {\it mixed composition \/} theoretical models at energies beyond $3\times 10^{18}$ eV does depend on the assumed HIM. Furthermore, at the highest energies, 
 $> 10^{19}$ eV, there could be an indication that the models give a systematically lighter composition than what the data suggests. Unfortunately, at present, statistics at these energies are too low for all experiments as to render any solid conclusion.

The most critical energy region for the understanding of the Galactic-extragalactic 
transition and disentangling its flux components, is $\sim 10^{17}$ - $\sim 3 \times 10^{18}$ eV. 
Inside this region, different experimental results, and they are different indeed, imply 
very different astrophysical scenarios, in particular with regard to the highest accelerators 
present in our own Galaxy. 

In the following sections we will center on two of the most important experimental results today, those of Auger and HiRes, 
and test what, if any, further modifications should be applied to the composition profile in order to make the data and the theoretical models as compatible as possible.

\subsection{Composition evolution along the transition region}

If our understanding of $\langle X_{max} \rangle$ in terms of composition is reasonably 
correct, Auger and HiRes elongation rate data suggest a composition 
profile along the ankle compatible with a mixed extragalactic composition, despite
differences in the energy dependence of $X_{max}$ for both experiments.

The agreement between Auger $X_{max}$ data \cite{Auger_Xmax_ICRC2007} and our 
predictions is remarkable at energies around $3\times10^{18}~eV$ for all HIMs. This
agreement extends to higher energies, even beyond $2 \times 10^{19}$ eV, for QGSJetI 
and QGSJetII v02 and v03, while EPOS and SIBYLL display different trends.
The picture is more complicated at lower energy where, under the present assumptions, there is no agreement regardless of the assumed HIM. 

The situation almost reverses in the case of the HiRes-MIA combined data set, for 
which a good agreement can be obtained at low energies with the EG {\it proton\/} 
model, but there is no clear fit at higher energies. 
 
It must be noted, however, that the composition profiles shown in Figs.\ref{EPOS_X_max_Allard_Bere}, 
\ref{QGSJet01_X_max_Allard_Bere}, \ref{QGSJetII_v02_X_max_Allard_Bere},  
 \ref{QGSJetII_v03_X_max_Allard_Bere} and \ref{Sibyll_X_max_Allard_Bere}, 
result simply from the combination of our previous solution to the total energy spectrum for 
the mixed model and the unmodified mixed model compositions as determined by 
\cite{Allard2005b}, plus fixed compositions for GA1 and GA2.

Therefore, in this section we analyze whether the composition profile of the
two additional Galactic components and of the extragalactic flux can be modified 
in a suitable way in order to satisfy, simultaneously, the existing 
spectral and elongation rate data along the transition region.

We use to this end a simplified model in which the shape of the 
extragalactic energy spectrum is kept, to first order, identical to Allard's mixed 
composition model 
\cite{Allard2007}, but the composition is limited to just two nuclei, proton and Iron, 
whose relative abundances can change appropriately in order to reproduce the behavior of
$\langle  X_{max} \rangle$ as a function of energy. The admitted lack of consistency in 
this approach is, we believe, compensated by the insight gained into the phenomenological 
constraints imposed by the data on the astrophysical models at play\begin{footnote}{An alternative analysis to the spectral fit is the use of slanted shower
data which, for Akeno data was performed by \cite{Stanev1993}.}\end{footnote}. 

On the other hand, the spectral shapes of both additional components and their 
normalizations are preserved, but their compositions as a function of energy are 
now functions to be determined during the fitting process. A binary mixture of 
p and Fe is also assumed. 

We further assume, a priori, that the diffusive Galactic spectrum and its composition 
are the ones determined previously in \S\ref{Combined spectrum} with the calculated 
renormalization of the CNO group. The diffusive Galactic flux from SNRs, 
spectrum as well as composition, is kept constant afterward during the fitting procedure.

We apply this procedure to both, Auger and HiRes data in order to gain an insight on how
present experimental uncertainties can affect our astrophysical understanding of the 
transition region.

\subsubsection{Auger data}\label{Auger_X_max_rep}

\begin{figure}
\begin{center}
\includegraphics [width=1\textwidth]{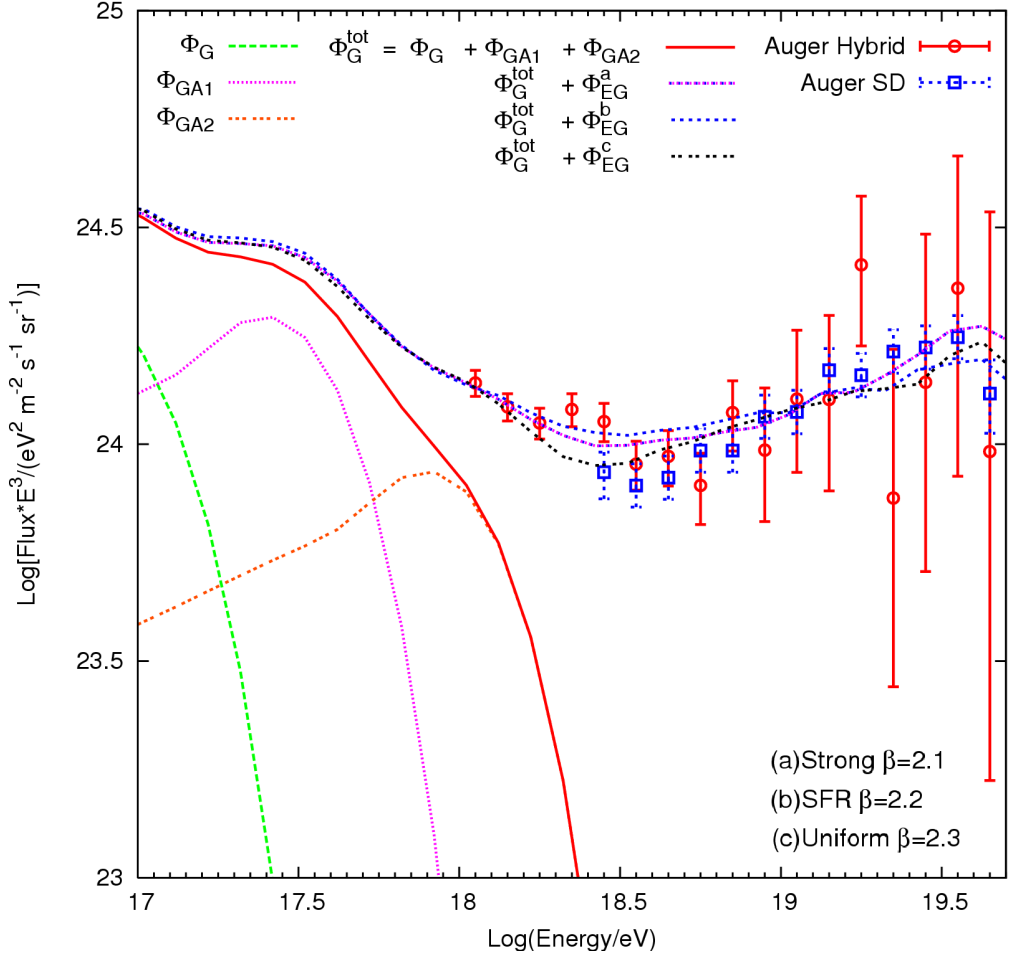}
\end{center}
\caption{Combined total spectrum $\Phi_G^{tot}+ \Phi_{EG}^{a,b,c}$ renormalized to Auger spectrum data for different EG evolution models (\S\ref{EG}). 
The Galactic spectrum $\Phi_G$, the Galactic additional components $\Phi_{GA1},~\Phi_{GA2}$ 
and  the total Galactic spectrum $\Phi_G^{tot}$ are also shown.
The normalization of the second Galactic additional component is the one determined for the EG ``uniform source distribution" model. See the text for details.} \label{Auger_spectrum}
\end{figure}

Fig.\ref{Auger_spectrum} shows the total spectrum and Galactic components 
that are used in this section. 
For each source evolution model (see \S\ref{EG}), 
we renormalize the EG spectrum to the surface 
detector (SD) Auger flux at $10^{19.05}$ eV \cite{AugerSD_ICRC2007}, where 
one can expect a negligible contribution from the Galactic CR flux. 
Afterward, we renormalize the GA1 and GA2 in order 
to match the total spectrum with the Auger spectrum data; the hybrid Auger 
spectrum\begin{footnote}{The Hybrid technique consists in reconstructing the CR showers 
using the complementary information given by the two independent detectors, the 
fluorescence detector (FD) and the surface detector (SD). These data allow the 
reconstruction of the cosmic ray energy spectrum up to energies 
below the mid-ankle.}\end{footnote} 
at lower energy, $10^{18}-10^{18.3}$ eV \cite{AugerFD_ICRC2007}, is used to this end.
The lowest energy branch of the ankle is approximated by a smooth interpolation 
between the highest energy KASCADE spectrum and the lowest energy hybrid Auger data.
 
It must be noted that, regardless of the renormalization, the extragalactic theoretical spectral 
models are too soft to allow a good fit to the Auger data at the highest energies. This 
inconsistency have also effects at lower
energies, where the position of the ankle is artificially pushed down from the position 
determined by Auger at $\sim 10^{18.6}$ eV. Therefore, as a compromise, we 
consider here the model corresponding to a ``uniform source distribution", 
which is the hardest one and provides the best visual fit to the highest energy data. 

\begin{figure}
\begin{center}
\subfigure[EG composition: proton (left) and Iron (right) components.]{
\includegraphics [width=1\textwidth]{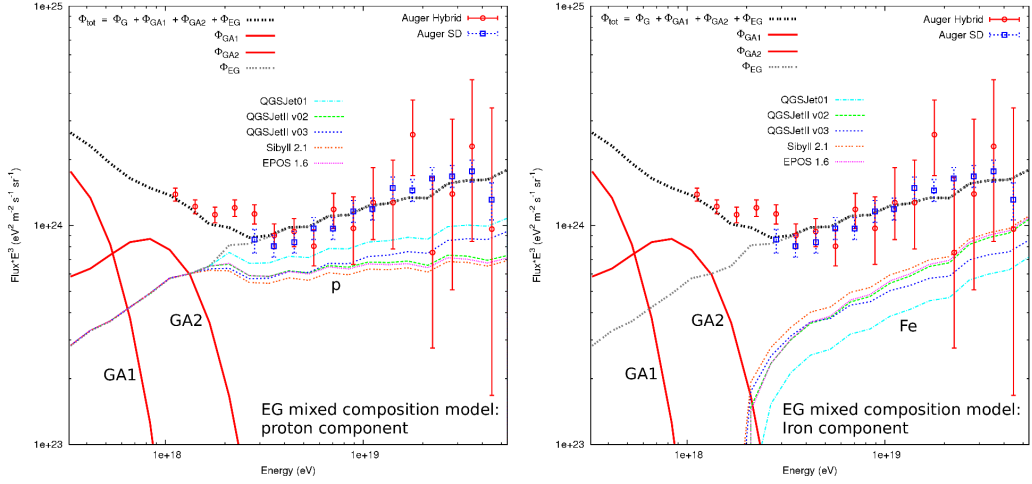}
\label{Auger_EG}
}
\subfigure[GA1 and GA2 composition: proton (left) and Iron (right) components.]{
\includegraphics [width=1\textwidth]{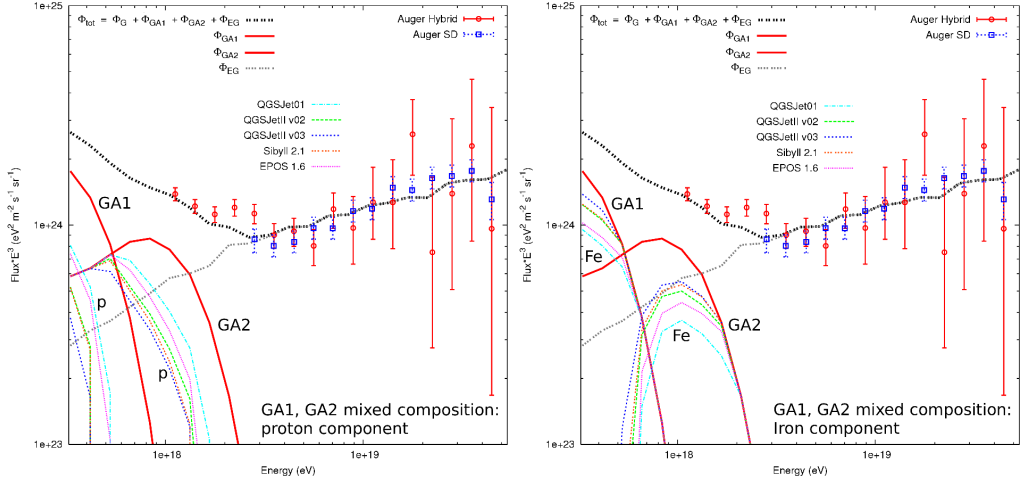}
\label{Auger_GA2}
}
\caption{Auger data. (a) EG proton and Iron composition;  
(b) GA1 and GA2 proton and Iron compositions. 
The fluxes are calculated, for each HIM, in order to reproduce the Auger $\langle X_{max}\rangle$ data. 
The  Galactic additional components ($\Phi_{GA1}$, $\Phi_{GA2}$), the EG spectrum ($\Phi_{EG}$)
and the total spectrum ($\Phi_{tot}=\Phi_{G}+\Phi_{GA1}+\Phi_{GA2}+\Phi_{EG}$) 
are also shown, as well as the Auger spectrum data.}\label{Auger_GA_EG}
\end{center}
\end{figure}

From the estimated combined Galactic-extragalactic total spectrum and 
using the $X_{max}$ parameterization given in \S\ref{Hadronic interaction models}, eq. \ref{eq:X_max}, 
we compute the relative abundances of proton and Fe nuclei of the EG spectrum and of the two additional Galactic components (GA1 and GA2)
that match  the Auger $\langle X_{max} \rangle$ data. The EG, GA1 and GA2 compositions are 
determined in the energy range of available Auger $X_{max}$ data, $\sim 3 \times 10^{17}$ to 
$\sim 5\times10^{19}$, for all the hadronic interaction models.

The corresponding Iron and  proton fluxes for the EG spectrum and  for GA1 and GA2 
are shown in Fig.\ref{Auger_GA_EG}. The different curves for each nucleus correspond 
to inferences obtained with different HIMs.

The corresponding average atomic weights, $<A>$, for each HIM are shown in Fig.\ref{Auger_HiRes_A}.

\subsubsection{HiRes data}\label{HiRes_X_max_rep}

\begin{figure}
\begin{center}
\includegraphics [width=1\textwidth]{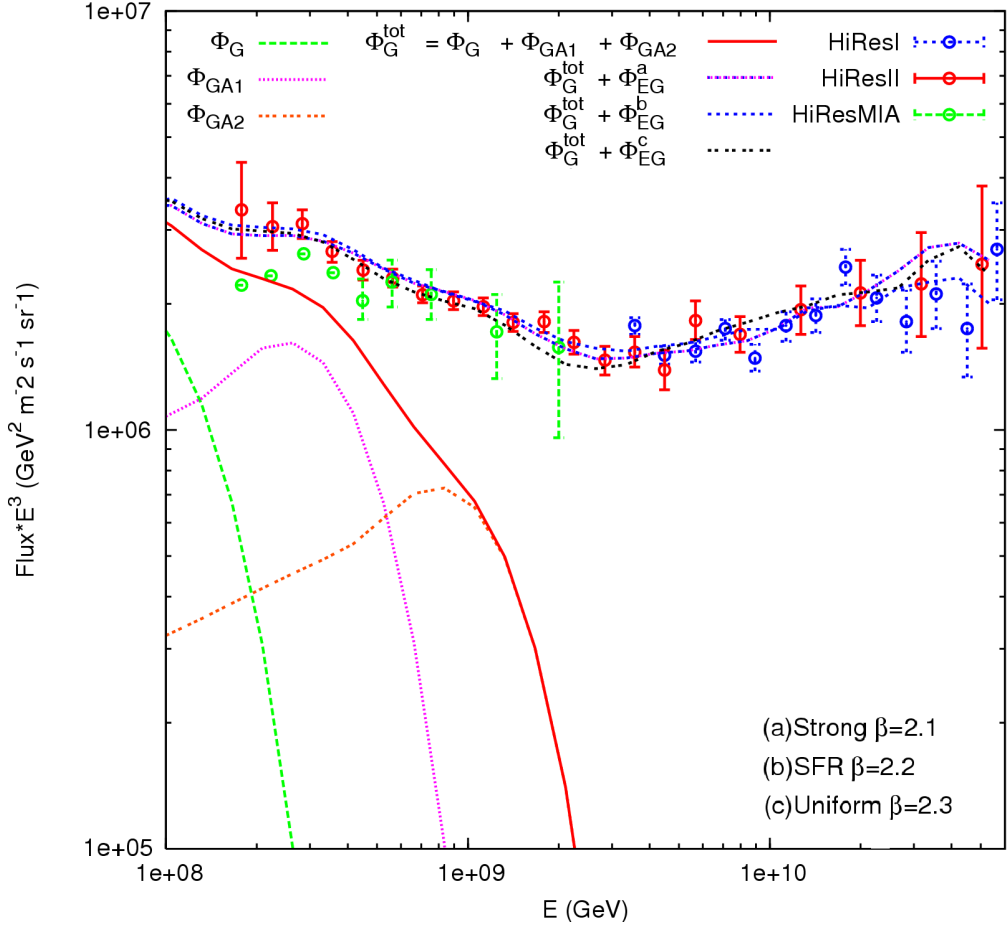}
\end{center}
\caption{Combined total spectrum $\Phi_G^{tot}+ \Phi_{EG}^{a,b,c}$ renormalized to HiRes spectrum data for different EG evolution models (\S\ref{EG}).
The Galactic spectrum $\Phi_G$, the Galactic additional components $\Phi_{GA1},~\Phi_{GA2}$
and  the total Galactic spectrum $\Phi_G^{tot}$ are also shown. 
The normalizations of the two Galactic additional component are the ones determined for the EG ``SFR" model.} \label{HiRes_spectrum}
\end{figure}

In order to assess the astrophysical implications of the present experimental uncertainties in the total
spectrum and elongation rate, we also apply here the same procedure as in the previous section to the HiRes data \cite{HiRes2005a,HiRes2005b,HiResMIA2001}. Fig. \ref{HiRes_spectrum} shows the
total spectrum, for different assumptions regarding the redshift evolution of the sources, 
and the Galactic components under consideration.
Since the HiRes spectrum extends to lower energies than those of Auger, in this case
we renormalize both additional Galactic components to match the HiResII data in the energy 
range $10^{17.3}-10^{18.3}$ eV (Fig.\ref{HiRes_spectrum}). The EG spectra, on the 
other hand, are already normalized to HiRes data (\S \ref{EG}).

As in the Auger case, the position of the ankle measured by HiRes (~$6\times10^{18}$ eV) is not accurately reproduced with any of the combined spectra if a reasonable agreement at higher energies is required. Consequently, we select for subsequent analysis the EG ``SFR" model, which is the one that provides the best visual fit to the experimental data. 

We assume, as in \S\ref{Auger_X_max_rep}, that the composition of the EG spectrum 
and of the two Galactic additional components is a binary mixture of proton and Iron 
nuclei and compute their relative abundances in order to match the 
$\langle X_{max} \rangle$ energy dependence of the HiRes data.

The compositions as a function of energy of EG, GA1 and GA2 are determined for 
all the HIM in the energy range of HiRes and HiResMIA $X_{max}$ data 
$\sim 10^{17}$ to $\sim 2 \times 10^{19}$.

\begin{figure}
\begin{center}
\subfigure[EG composition: proton (left) and Iron (right) components.]{
\includegraphics [width=1\textwidth]{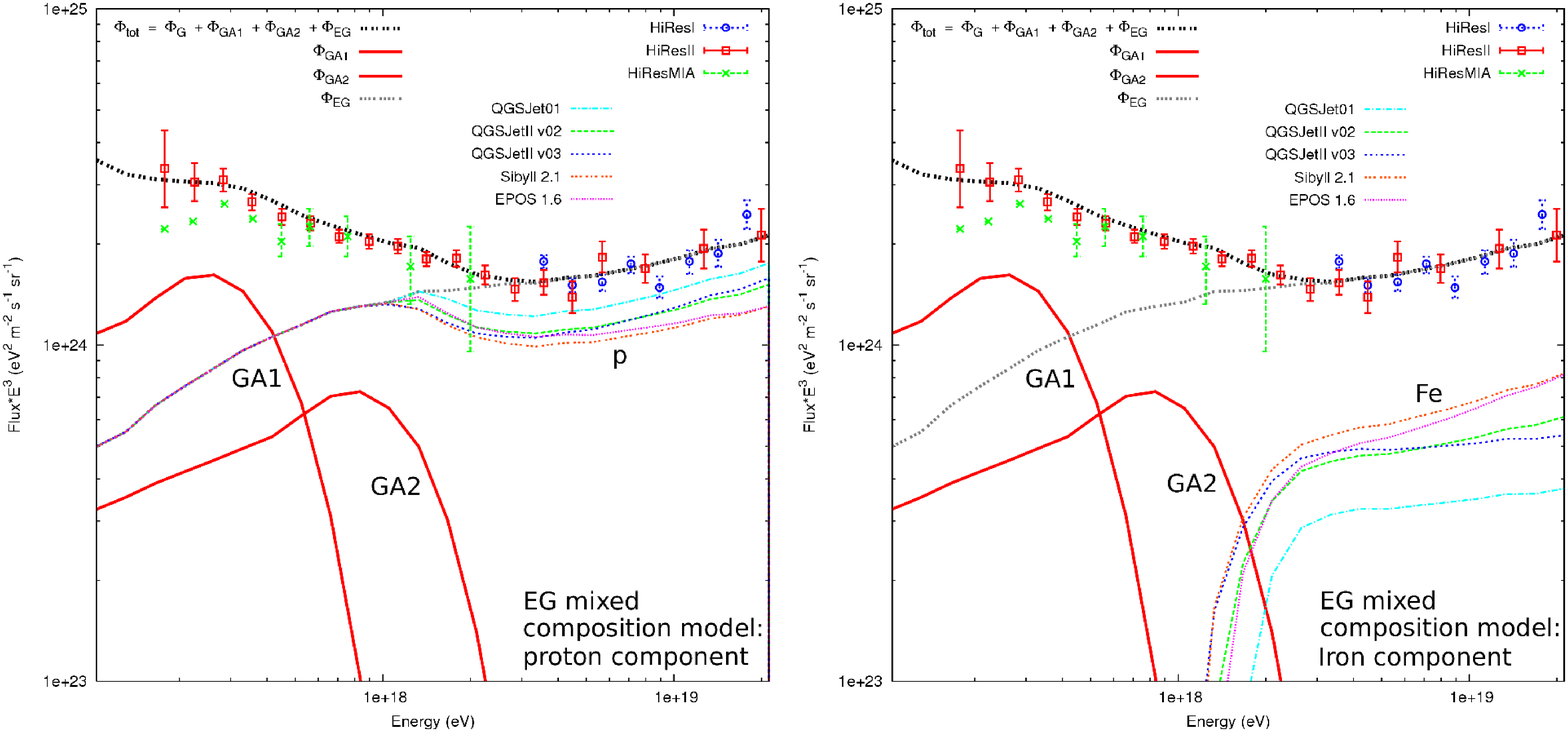}
\label{HiRes_EG}
}
\subfigure[GA1 and GA2 composition: proton (left) and Iron (right) components.]{
\includegraphics [width=1\textwidth]{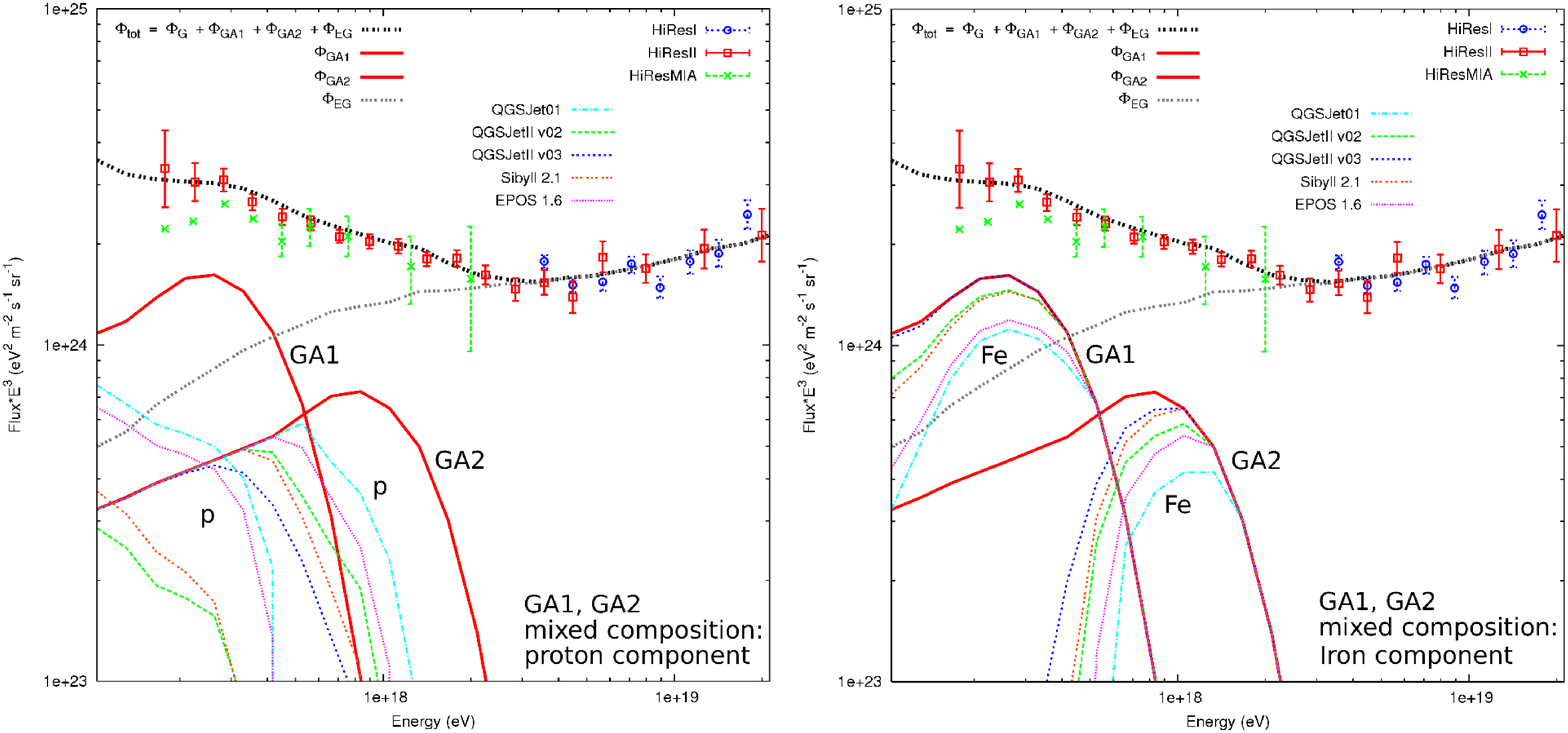}
\label{HiRes_GA2}
}
\caption{HiRes data. (a) EG  proton and Iron composition; (b) GA1 and GA2 proton and Iron compositions.
The fluxes are calculated for each HIM in order to reproduce the  HiRes and HiResMIA $\langle X_{max}\rangle$ data.
The  Galactic additional components ($\Phi_{GA1}$, $\Phi_{GA2}$), the EG spectrum ($\Phi_{EG}$)
and the total spectrum ($\Phi_{tot}=\Phi_{G}+\Phi_{GA1}+\Phi_{GA2}+\Phi_{EG}$)
are also shown,  as well as the HiRes spectrum data.}\label{HiRes_GA_EG}
\end{center}
\end{figure}

The results for the calculated Iron and proton fluxes for GA1, GA2 and EG are shown in 
Fig.\ref{HiRes_GA_EG}.

\begin{figure}
\begin{center}
\includegraphics [width=1\textwidth]{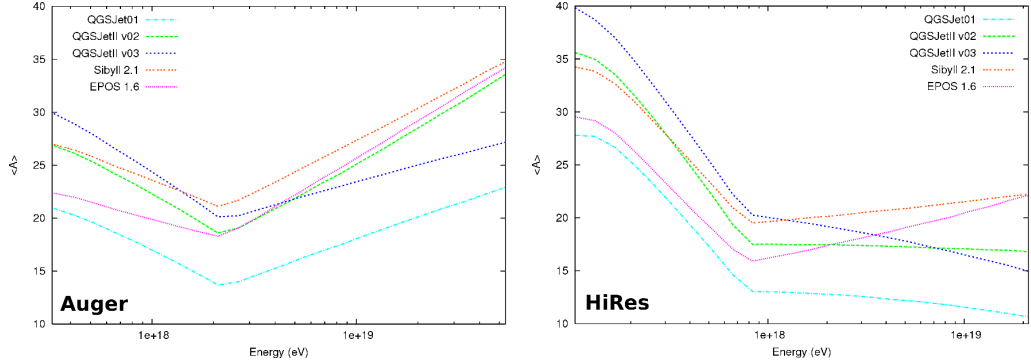}
\end{center}
\caption{Average atomic weight of the combined cosmic ray flux: the values were obtained 
from the calculated EG, GA1 and GA2 compositions in order to reproduce the Auger 
(Fig.\ref{Auger_GA_EG}) and HiRes (Fig.\ref{HiRes_GA_EG}) $\langle X_{max}\rangle$ data, 
under different HIM assumptions. }\label{Auger_HiRes_A}
\end{figure}

The average atomic weights, $\langle A \rangle$, are shown in Fig.\ref{Auger_HiRes_A}, for each 
HIM.

\subsection{Discussion on composition}

In section \ref{XmaxProfiles} we calculated the $X_{max}$ dependence on energy 
for all the possible scenarios obtained by combining the calculated SNR Galactic spectrum with different extragalactic models (\S\ref{Combined spectrum}). Several HIMs were used.

Simple assumptions were made on the composition of the two additional Galactic components.
The first one, which is required by both EG models, was assumed to be Iron dominated, as it is probably contributed by compact and highly magnetized SNRs. The origin and properties of GA2, required in the case of a mixed composition EG spectrum, 
are more uncertain.  Consequently, as an exploratory test, we considered in section 
\ref{XmaxProfiles} two cases for GA2: a pure proton and a pure Iron composition.

For all the possible combined spectra, we used different parameterizations of 
$\langle X_{max}\rangle$ obtained for the HIMs used in literature to calculate the behavior of $X_{max}$ in the energy range $[10^{17}~eV,10^{20}~eV]$.
We compared the calculated $X_{max}$ energy profiles with the available experimental data from Fly's Eye, HiRes, HiRes-MIA and Auger. 

The EG {\it proton\/} model is very much compatible with HiRes-MIA data at energies below
$10^{18}$ eV for any HIM. However, if the current interpretation of $X_{max}$ in terms of shower composition is correct, the later model is completely incompatible with higher energy data which
 points to a mixed composition.

In the case of the EG {\it mixed-composition\/} model, also two energy regimes are clearly distinguishable. 

At energies above $\sim 3 \times 10^{18}$ the slope of the elongation rate
seems to be consistent with that of the data up to, possibly, $\sim 2 \times 10^{19}$ eV.
The effect of different HIMs is basically to change the normalization of the curves, changing at the same time the experimental data set with which the models are more compatible. For energies 
smaller than $\sim 3 \times 10^{18}$ eV, the calculated $\langle X_{max} \rangle$ profiles are 
very dependent on the assumption made regarding the composition of GA2. A heavy GA2
can fit reasonably well the Fly's Eye data, while a light GA1 is more consistent with HiRes-MIA
in all the energy interval and with Auger between $4-6 \times 10^{17}$ eV. Between $\sim 6 \times 10^{17}$ and $\sim 2 \times 10^{18}$ eV Auger data is more or less in-between both solutions.
These results are qualitatively independent of the HIM. Even if the assumption of a pure chemical composition for GA1 and GA2 is artificial, it is clear that the present discrepancies between 
the various experimental results are of astrophysical significance, since they have quite 
different implications with respect to the nature of most powerful Galactic accelerators. 
However, and surprisingly enough, the uncertainty associated with the several HIM currently in 
use in the literature does not pose a too large qualitative problem from the astrophysical point 
of view; that is, unless our present understanding of the hadronic interaction processes is not, somehow, fundamentally wrong.

In sections \S\ref{Auger_X_max_rep} and \S\ref{HiRes_X_max_rep} we 
determined the composition, as a function of energy, of the extragalactic 
spectrum (EG) and the two additional Galactic components (GA1, GA2) that 
fits $X_{max}$ data of Auger and HiRes, under the simplifying assumption of a 
binary mixture of p and Fe for the cosmic ray flux.

The main result is that, regardless of the experimental data set considered, the 
composition has to be mixed to some extent along all the spectrum. The assumed HIM
plays a very important quantitative role in the inferred composition of any individual
component.

An important point to note regarding GA1 and GA2 is that, even if they have mixed 
composition, the composition profile inside each one of them is similar in the sense that
the individual fluxes are lighter at lower energies and then become heavier as the energy
increases. This is a systematic effect that is quantitatively, but not qualitatively affected
by the HIM used. Profiles such of those we obtained for GA1 and GA2 are similar to
what can be expected from different populations of SNRs immersed in differing 
environments. If this is correct, then probably only SNRs are required in order to explain 
the main part of the Galactic flux up to energies $\lesssim 3 \times 10^{18}$ eV. In any
case, GA2 is lighter than GA1, which could be an evidence for the existence of a minor
contribution coming from a different source like, for example, inductors associated 
with compact objects.

The estimated average atomic weight (see, Fig. \ref{Auger_HiRes_A}) varies widely 
depending on the assumed HIM. However, it can be seen that, for a given HIM, HiRes
always implies heavier compositions on the Galactic side of the transition region than
Auger. Despite this systematic difference, both data sets show the same trend of
a decreasing value of $\langle A \rangle$ as a function of energy.

The opposite happens at higher energies on the extragalactic side of the flux,
where Auger implies an increasing $\langle A \rangle$ as a function of energy regardless 
of the HIM, and HiRes predicts either a rather constant or a much more slowly changing 
composition. Also, for the latter data set, the uncertainties introduced by the HIM are 
qualitatively more important, since the high energy slope of the profile can change 
signs depending on the adopted interaction model.

It must be noted that, for both data sets, there seems to be a break in the average 
energy composition profile at an energy around $\sim 10^{18}$ eV or slightly higher,
which may further highlight the physical association between the ankle feature in the 
cosmic ray spectrum an the Galactic-extragalactic flux transition.

\section{Conclusions}

We have analyzed the matching conditions of the Galactic and
extragalactic components of cosmic rays along the second knee and
the ankle.

We have calculated the diffusive Galactic spectrum from regular SNRs 
using the numerical diffusive propagation code GALPROP. We used this 
first assessment of the Galactic cosmic ray flux to analyze the matching
conditions with two alternative models for the extragalactic flux: a pure
proton model \cite{Berezinsky2006c} and a mixed composition model 
\cite{Allard2007}.

The first step was the matching of the observed energy spectrum by
combining the Galactic and extragalactic fluxes. From this process, it 
becomes clear that additional Galactic components must be at play in
reality. The minimum amount of additional components that allows us
to satisfactorily reproduce the spectrum is either one or two. The pure {\it proton}
model is the less expensive one with only one required additional component, GA1. 
The {\it mixed} composition model on the other hand requires, besides GA1, 
another components at still higher energies, GA2.

If no other information is used, both theoretical models are indistinguishable
from the experimental point of view. Therefore, we also analyzed the effect 
of incorporating composition information, in the form of elongation rate data
($\langle X_{max} \rangle$). For this study, the shape of each one of the
components is determined from the matching of the observed total energy
spectrum. The elongation rate of the combined fluxes is then fitted to the 
HiRes and Auger data sets by changing appropriately the composition as
a function of energy of each one of the components, Galactic and 
extragalactic. In order to make the  analysis simpler, only a binary mixture
of p and Fe is considered. The main result is that the additional Galactic components,
GA1 and GA2, must have a mixed composition. Furthermore, inside each one of 
this components there is a progressive evolution of the composition from lighter
to heavier as the energy increases. This is consistent with this components 
being originated in different populations of SNRs. Additionally, GA2 is  
globally lighter than GA1, which could indicate the possible existence of
a minor contribution from another acceleration mechanisms, without a rigidity 
cut-off, operating at the highest energies. 

The uncertainties introduced by the HIM are always important from the quantitative
point of view. From the qualitative point of view, however, there are some 
results that seem rather independent of the HIM, like the diminution in average
atomic mass along the low energy branch of the ankle and the existence of a
discontinuity in the slope of the energy profile of $\langle A \rangle$ around 
the mid ankle. These results seem also to be supported by both, Auger and
HiRes.

Our present results may be certainly considered as preliminary due to 
experimental uncertainties and simplifications in the numerical approach. 
First, there is the paucity of data involved in the determination of the 
energy spectrum in the region encompassing the second knee and the ankle 
and the divergence between $\langle X_{max} \rangle$ measurements by
different experiments below $3 \times 10^{18}$ eV. A proper experimental 
characterization of this very important region will likely have to wait until the release
of the KASCADE-Grande \cite{KASCADE-Grande2008} and Auger enhancement data \cite{GMTanco2007}. 
Second, there are arguable simplifying assumptions related with our diffusive 
treatment of the Galactic component at the highest energies which is, very 
likely, undergoing a change in propagation to a full ballistic regime. This can be 
somehow mitigated by the fact that Fe nuclei should still be diffusive inside this
energy interval, while protons would only deviate importantly from the diffusive
approximation at the highest energies considered for the Galactic flux.
In any case, the importance of the transition region as a play ground for 
disentangling the Galactic and extragalactic cosmic ray fluxes is unquestionable
and considerable effort should be invested in its full experimental characterization 
and theoretical modeling.

\section*{Acknowledgments}

CDD is partially supported by PhD grant from the Universit\`a degli Studi di Milano.
CDD also thanks the Inst. de Ciencias Nucleares, UNAM, for its hospitality during several extended stays. GMT thanks CONACyT and PAPIIT/CIC-UNAM, through grants IN115707
and IN115607, for partial support. Both authors have benefitted from their interaction with
several colleagues from the Pierre Auger Collaboration.

\bibliographystyle{elsart-num}

\bibliography{DeDonato_MedinaTanco_bibliography}









\end{document}